\documentclass[%
 aip,
 amsmath,amssymb,
 reprint,%
]{revtex4-1}

\usepackage{graphicx, lipsum}
\usepackage{dcolumn}
\usepackage{bm}
\usepackage{gensymb} 
\usepackage{colortbl} 
\usepackage{array}
\usepackage{makecell} 
\usepackage{tabularx, booktabs} 
\usepackage[utf8]{inputenc}
\usepackage[T1]{fontenc}
\usepackage{mathptmx}
\usepackage{etoolbox}
\usepackage{textcomp}
\usepackage{multirow}
\usepackage{hhline}
\usepackage{caption}
\usepackage[justification=raggedright]{caption}
\usepackage{subcaption}
\usepackage{wrapfig}
\usepackage{graphicx}
\usepackage[dvipsnames]{xcolor}
\usepackage{float}
\usepackage{microtype}
\definecolor{Gray}{gray}{0.9}
\newcommand{\RomanNumeral}[1]{%
  \uppercase\expandafter{\romannumeral #1}%
}
\makeatletter
\def\@email#1#2{%
 \endgroup
 \patchcmd{\titleblock@produce}
  {\frontmatter@RRAPformat}
  {\frontmatter@RRAPformat{\produce@RRAP{*#1\href{mailto:#2}{#2}}}\frontmatter@RRAPformat}
  {}{}
}%
\makeatother

\begin{document}

\title{Fabrication and Structural Analysis of Trilayers for Tantalum Josephson Junctions with Ta$_2$O$_5$ Barriers}

\author{Raahul Potluri}
\affiliation{
Department of Electrical \& Computer Engineering, University of Washington, Seattle, WA 98195, USA
}

\author{Rohin Tangirala}
\affiliation{
Department of Electrical \& Computer Engineering, University of Washington, Seattle, WA 98195, USA
}

\author{Jiangteng Liu}
\affiliation{
Department of Electrical \& Computer Engineering, University of Washington, Seattle, WA 98195, USA
}

\author{Alejandro Barrios}
\affiliation{Shared Instrumentation Facility, Colorado School of Mines, Golden, CO 80401, USA}

\author{Praveen Kumar}
\affiliation{Shared Instrumentation Facility, Colorado School of Mines, Golden, CO 80401, USA}

\author{Sage R. Bauers}
\affiliation{Materials Science Center, National Laboratory of the Rockies, Golden, CO, 80401 USA}

\author{Peter V. Sushko}
\affiliation{Physical and Computational Sciences Directorate, Pacific Northwest National Laboratory, Richland, WA 99354, USA}

\author{David P. Pappas}
\affiliation{Rigetti Computing, Inc., Berkeley, CA 94710, USA}

\author{Serena Eley}
\email{serename@uw.edu}
\affiliation{
Department of Electrical \& Computer Engineering, University of Washington, Seattle, WA 98195, USA
}

\date{\today}

\begin{abstract}

Tantalum (Ta) has emerged as a promising low-loss material, enabling record coherence times in superconducting qubits. This enhanced performance is largely attributed to its stable native oxide, which may host fewer two-level system (TLS) defects --- key contributors to decoherence in superconducting circuits. Nevertheless, aluminum oxide remains the predominant choice for Josephson junction (JJ) barriers in most qubit architectures. Here, we investigate techniques for forming high-quality oxide layers on $\alpha$-phase tantalum films to develop tantalum-oxide JJ barriers. We explore thermal oxidation in a tube furnace, rapid thermal annealing, and plasma oxidation of both room-temperature and heated Ta films, characterize the resulting structures using X-ray techniques and electron microscopy, and propose a mechanistic picture of the oxidation pathways. We find that plasma oxidation provides the smoothest Ta$_2$O$_5$ layers, is compatible with in situ Ta deposition, and offers thickness control through the annealing temperature, advantageous for JJ fabrication. Lastly, we evaluate methods for growing Ta/TaO$_x$/Ta trilayers. All trilayers showed c‑axis‑oriented columnar growth of the bottom Ta layer, with sapphire substrates producing larger, better‑aligned grains yet higher dislocation densities than silicon. Nucleation of c‑axis‑oriented $\alpha$-Ta on tantalum-oxide required an Nb seed layer, as direct Ta deposition yielded amorphous Ta. These results demonstrate the feasibility of $\alpha$-Ta/Nb/TaO$_x$/$\alpha$-Ta stacks for JJs with clean interfaces.

\end{abstract}

\maketitle

\section{\label{sec:intro}Introduction\protect}

Superconducting circuits designed to operate as artificial atoms, exhibiting discrete energy levels, are a promising platform for scalable quantum computing. These micron-sized circuits combine inductive and capacitive components with nanoscale Josephson junctions --- two superconductors sandwiching a non-superconducting barrier layer. Josephson junctions induce a non-linear response such that the two lowest energy states can be selectively addressed. Information is then encoded in the charge, flux, or phase of the circuit and stored for time frames limited by decoherence.\cite{Oliver2013} Although the precise microscopic origins of decoherence in Josephson junctions are not completely understood, they can be traced to defects in the superconductor, the tunneling barrier, and at the interfaces. These defects become problematic when they form two‑level systems (TLSs) with electric dipole moments that couple to operating currents, leading to energy relaxation and dephasing.\cite{Phillips1987,Muller2019,Eckstein2013,deLeon2021}

Aluminum is often considered the silicon of superconducting qubits. Most fabrication processes take advantage of aluminum's self-terminating native oxide to create an amorphous tunnel barrier in aluminum-based Josephson junctions.\cite{Lang_2023} However, this native oxide is known to be a significant source of defects that contribute to decoherence, leading researchers to explore alternative materials.\cite{deLeon2021,Zeng_2015,Martinis2005,Lisenfeld2015,Lisenfeld2019} Notably, tantalum (Ta) has recently emerged as a promising material for low-loss quantum circuits. Tantalum resonators have demonstrated high quality factors up to $15 \times 10^6$ in the single-photon regime,\cite{DeLeon2023, Urade2024, Shi2022} exceeding the highest values observed in aluminum resonators.\cite{Zikiy2023} Moreover, tantalum-based qubits have achieved record lifetimes ($T_1$) of above 0.5 ms,\cite{Place2021, Wang2022} a significant improvement over aluminum-based qubits.\cite{Keller2017} However, these Ta-based qubits still depend on lossy AlO\(_x\) junction barriers.

The native oxide of tantalum (TaO$_x$) is thinner, forms cleaner interfaces,\cite{McLellan2023} and may contain fewer TLS defects than that of niobium (Nb), another commonly-used material in superconducting circuits.\cite{Wang2025} In particular, the tantalum's native oxide exhibits a closer-to-crystalline bonding nature with less disorder than that of Nb, likely mitigating the creation of TLSs by limiting atomic hydrogen diffusion to the interface.\cite{Oh2024} The stability of the oxide has also been demonstrated through high quality factors in tantalum resonators measured over several months.\cite{Ding2024} Given these favorable characteristics, it is compelling to investigate whether tantalum-based Josephson junctions—featuring tantalum-oxide barriers—can achieve reduced losses.  An advantage of oxidizing the superconductor to form a barrier is that it can yield highly conformal, pinhole‑free barriers, a level of coverage that is difficult to obtain in directly grown, few-nm-thick barriers. This exploration necessitates establishing a reliable growth method for tantalum-oxide and trilayer structures in which superconducting $\alpha$-phase tantalum encloses the tantalum-oxide barrier.

\begin{table*}[t!]
\caption{\label{tab:table1}Summary of sample properties and results, including the substrate, Ta thickness ($d_{Ta}$), oxygenation process temperature (Temp.), oxygen flow rate, oxygen annealing time, X-ray reflectometry, and atomic force microscopy results for oxidized samples. All samples on Si substrates include a sputtered 6-nm Nb seed layer below the Ta layer. Plasma oxidation was performed with 100 W of RF power biasing the substrate.}

\begin{ruledtabular}
\begin{tabular}{ccccccccc}
 &  & & \multicolumn{4}{c}{\cellcolor{gray!20}\textbf{Oxygenation Process}} & & \\
\textbf{ID} & \textbf{Substrate} & \textbf{$d_{\text{Ta}}$} & \textbf{Method} & \textbf{Temp.} & \textbf{O$_2$} & \textbf{Time} & \textbf{TaO$_x$ thickness} & \textbf{AFM roughness} \\
 &  & \textbf{[nm]} &  & \textbf{[°C]} & \textbf{[sccm]} & \textbf{[min]} & \textbf{[nm]} & \textbf{[nm]} \\
\hline
1 & Si & 100 & --- & 22 & --- & --- & 1.93 & 0.518\\
2 & Al$_2$O$_3$ & 150 & --- & 22 & --- & --- & 2.38 & 0.72\\
\hline
3 & & & & & & 10 & 37.14 & 1.04\\
4 & Si & 100 & Tube Furnace & 400 & 20,000 & 30 & 56.66 & 1.24\\
5 & & & & & & 60 & 60.32 & 1.71\\
\hline
6 & & & & & & 1 & --- & ---\\
7 & Si & 100 & RTA & 700 & 5,000 & 5 & --- & ---\\
8 & & & & & & 10 & --- & ---\\
\hline
9 & & 60 & & 21.3 & & 10 & 7.64 & 0.572\\
10 & & 60 & & 21.3 & & 30 & 7.33 & 0.613\\
11 & & 100 & & 200 & & 60 & 7.46 & 0.108\\
12 & Si & 100 & Plasma & 200 & 20 & 120 & 7.78 & 0.105\\
13 & & 100 & & 300 & & 60 & 10.35 & 0.463\\
14 & & 100 & & 300 & & 120 & 10.78 & 0.125\\
15 & & 100 & & 400 & & 60 & 15.21 & 0.342\\
16 & & 100 & & 400 & & 120 & 14.88 & 0.232\\
\end{tabular}
\end{ruledtabular}

\end{table*}

In this study, we test three protocols for creating a tantalum-oxide layer for Ta/TaO$_x$/Ta Josephson junctions --- tube furnace annealing, rapid thermal annealing, and oxygen plasma annealing in a sputtering system --- in each case varying the annealing temperature and time. We then measure the oxide thicknesses using X-ray reflectometry (XRR), characterize oxide species using X-ray photoelectron spectroscopy (XPS), analyze grain structure using atomic force microscopy (AFM) and electron microscopy, and construct the mechanistic picture of the oxidation process using ab initio modeling based on the density functional theory (DFT). Lastly, we test different processes for growing $\alpha$-phase Ta on top of the oxidized Ta layer, to form a trilayer structure necessary for a Ta-based Josephson junction.

\section{\label{sec:experimentaldetails}Experimental Details\protect}

Crystalline Ta exists in two phases: the alpha phase ($\alpha$-Ta), which has a body-centered cubic (BCC) structure, and the beta phase ($\beta$-Ta), which has a tetragonal crystalline structure and is metastable.\cite{Colin2017, Clevenger1992} Because $\alpha$-Ta has a superconducting critical temperature $T_c$ that is substantially higher than that of $\beta$-Ta (4.4 K versus 0.6-1 K), $\alpha$-Ta is preferred for superconducting circuit applications. 

To promote the growth of $\alpha$-Ta, we deposited Ta via DC magnetron sputtering using two different procedures: high-temperature growth on sapphire substrates and room-temperature growth on silicon substrates coated with a niobium (Nb) seed layer.\cite{Face1987,Alegria2023, Marcaud2025} In the former, 150 nm of Ta was deposited on sapphire substrates held at 500 $\degree$C, as substrate heating above 400 $\degree$C is known to result in the growth of pure $\alpha$-Ta.\cite{Gladczuk2004} In the latter, 60 or 100 nm (see Table \ref{tab:table1}) of Ta was deposited on Si(004) substrates atop a 6-nm Nb seed layer. Further details regarding the growth process are included in the Methods section. Lastly, we verified that we obtained $\alpha$-Ta using X-ray diffraction and measurements of a $T_c \approx$ 4.3 K using magnetometry, as shown in Supplementary Information Fig.~S1.

To develop a protocol for creating a tantalum-oxide barrier for fully tantalum-based Josephson junctions, we tested three different oxidation procedures: annealing Ta films in (1) an Expertech CTR200 tube furnace under a flow of oxygen, (2) an oxygen flow inside a AccuThermo AW 610 rapid thermal processing system, and (3) an oxygen plasma in a Lesker Lab 18 sputtering chamber. Table \ref{tab:table1} summarizes the procedures applied to each sample.

First, the tube furnace annealing process involved heating the system to 350$\degree$C at a rate of 10 $\degree$C/min then at 2.5 $\degree$C/min until the final temperature reached 400$\degree$C. To prevent any oxidation during the ramp-up process, the furnace was flushed with nitrogen (N$_2$) gas. Upon achieving 400$\degree$C, the system was flushed with O$_2$ at a rate of 20,000 sccm for 10-60 minutes (see Table \ref{tab:table1}). Immediately following the annealing process, the heater was turned off, and the system was allowed to cool under a continuous flow of nitrogen gas.

Second, to test a faster oxidation process, we employed rapid thermal annealing (RTA), taking advantage of its swift temperature ramp-up and ramp-down capabilities. The films were heated to 700$\degree$C at a rate of 70$\degree$C/sec under a 5,000 sccm flow of O$_2$ and held for durations ranging from 1 to 10 minutes, as detailed in Table \ref{tab:table1}. As with the tube furnace annealing method, the system was purged with N$_2$ during all other stages and subsequently cooled to room temperature following the annealing step.

Finally, for plasma oxidation, the films were exposed to a 100 W RF plasma in a sputtering chamber, using a 20 sccm flow of oxygen gas. The base pressure of the chamber prior to oxidation was less than $6 \times 10^{-8}$ torr. Oxidation was then carried out at room temperature, as well as with the substrate heated to 200$\degree$C, 300$\degree$C, and 400$\degree$C to assess temperature-dependent effects.

\section{\label{sec:results}Results and Discussion\protect}

When a non-superconducting barrier separates two superconducting electrodes, the superconductors' order parameters decay exponentially into the barrier. For a barrier thickness $d$ that is sufficiently thin compared to the superconductor coherence length ($\xi$), the order parameters overlap. This overlap enables phase coherence between superconductors and a supercurrent of superconducting electron pairs (Cooper pairs) to flow across the barrier. In a superconductor-insulator-superconductor (SIS) junction, Cooper pair tunneling dominates when the barrier thickness $d < \xi$, whereas quasiparticle tunneling dominates for thicker barriers, resulting in lower supercurrent.\cite{Poole2014}

In epitaxial Ta films that exhibit 2D superconductivity, the in-plane coherence length of $\alpha$-Ta has been measured to vary between 21.7 nm and 54.8 nm, for films up to 162.2 nm thick.\cite{Chen2025} These values set an effective upper limit to the thickness of an insulating barrier for SIS tunnel junctions based on Ta.  Consequently, we must develop a process that can produce thin, fully oxidized barriers without significant inter-metallic Ta that could form a superconducting short between electrodes. Moreover, it is important to identify the type of oxide formed for consideration of the potential sources of energy loss.

\subsection{\label{sec:XPS}Oxygen Species}

To determine the chemical state of the oxidized layer in our films, we performed X-ray photoelectron spectroscopy, a technique that employs the photoelectric effect to obtain the chemical composition of the top layer of a material. We used a Kratos Axis-Ultra DLD spectrometer equipped with an Al K$\alpha$ X-ray source operating at 15 kV and 225 W. For these acquisitions, the X-ray spot size was approximately 700 $\times$ 300 \textmu{}m$^2$. Note that this spectrometer has a sample probe depth of 10 nm.

\begin{figure}[htp!]
    \centering
    \includegraphics[width=1\linewidth]{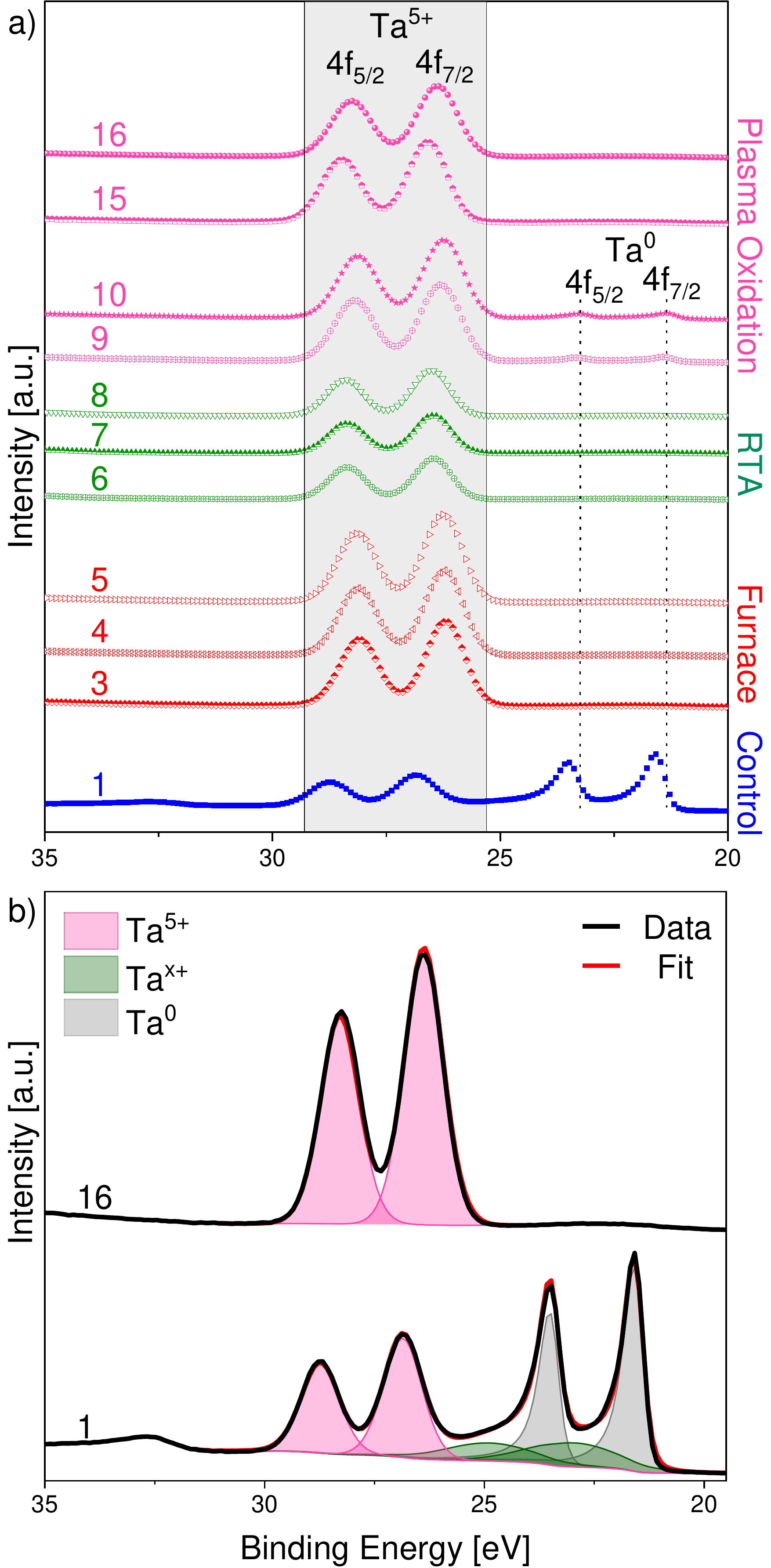}
    \caption{a) High-resolution Ta(4f) XPS spectra of tantalum films -- featuring a native oxide, tube furnace oxidation, rapid thermal annealing, and plasma oxidation samples showing the characteristic Ta$_2$O$_5$ doublet peaks at binding energies of approximately 26.2 eV (Ta 4f$_{7/2}$) and 28.1 eV (Ta 4f$_{5/2}$). For samples 1, 9, 10, peaks corresponding to metallic Ta appear near 21.4 and 23.3 eV due to XPS probing deeper than the oxide thickness. b) Peak-fit analysis of sample 1, 16 showing the different fits used to analyze the data. Dotted lines are a guide for the eye marking the positions of the metallic Ta.}
    \label{fig:XPS}
\end{figure}

Fig.~\ref{fig:XPS} shows the high-resolution photoelectron Ta(4f) spectra of the films. For the control film (sample 1), which contains only a native surface oxide, we resolve two pairs of double peaks originating from three spin–orbit–split doublets, with fits shown in Fig.~\ref{fig:XPS}(b). The doublet at 26.88 eV and 28.78 eV (pink) corresponds to the 4f$_{7/2}$ and 4f$_{5/2}$ valence states of Ta$^{5+}$ in Ta$_2$O$_5$.\cite{McGuire1973} The gray peaks at 21.58 eV and 23.48 eV similarly arise from 4f$_{7/2}$ and 4f$_{5/2}$, but in metallic Ta$^0$.\cite{McGuire1973} A final, broader doublet at 23.08 eV and 24.98 eV (yellow), with a full width at half maximum of 2.41 eV, corresponds to a mixed oxidation state of tantalum.\cite{Simpson2017}  These results are consistent with other XPS studies on Ta.\cite{Hu2023, Place2021, Wu1993, Chen1997, Muto1994} 

We observe a similar peak structure, but with only Ta$^{5+}$ and Ta$^{0}$, in the spectra collected from the two Ta films exposed to an oxygen plasma at room temperature (samples 9, 10). The spectra produced by the remaining films exhibit only the Ta$_2$O$_5$ doublet, suggesting complete oxidation within the top 10 nm of the film, results that are verified through oxide thickness measurements presented in Section \ref{sec:XRR}. For further composition analysis and the complete XPS spectra of these samples please refer to Supplementary Information section S2, including Supplementary Figs. S2 and S3 as well as Table S1 for a summary.

Comparing the Ta$^{5+}$ 4f XPS peaks between samples reveals small shifts in the binding‑energy (0.7–0.8 eV), consistent with temperature‑dependent oxygen‑vacancy formation. Different annealing temperatures create varying concentrations of oxygen vacancies in Ta$_2$O$_5$, with the formation of surface oxygen vacancies resulting in altered local electronic environments,\cite{Liu2020} while oxygen gas availability during processing contributes to reducing oxygen vacancy density.\cite{Lau2003} These process-induced variations in defect density create Ta$_2$O$_5$ layers with different electrical properties, leading to different degrees of surface charge buildup during XPS measurement, which shifts photoelectron binding energies.\cite{Baer2020}

\subsection{\label{sec:XRR}Oxide Thickness}

\begin{figure}[hbt!]
    \centering
    \includegraphics[width=1\linewidth]{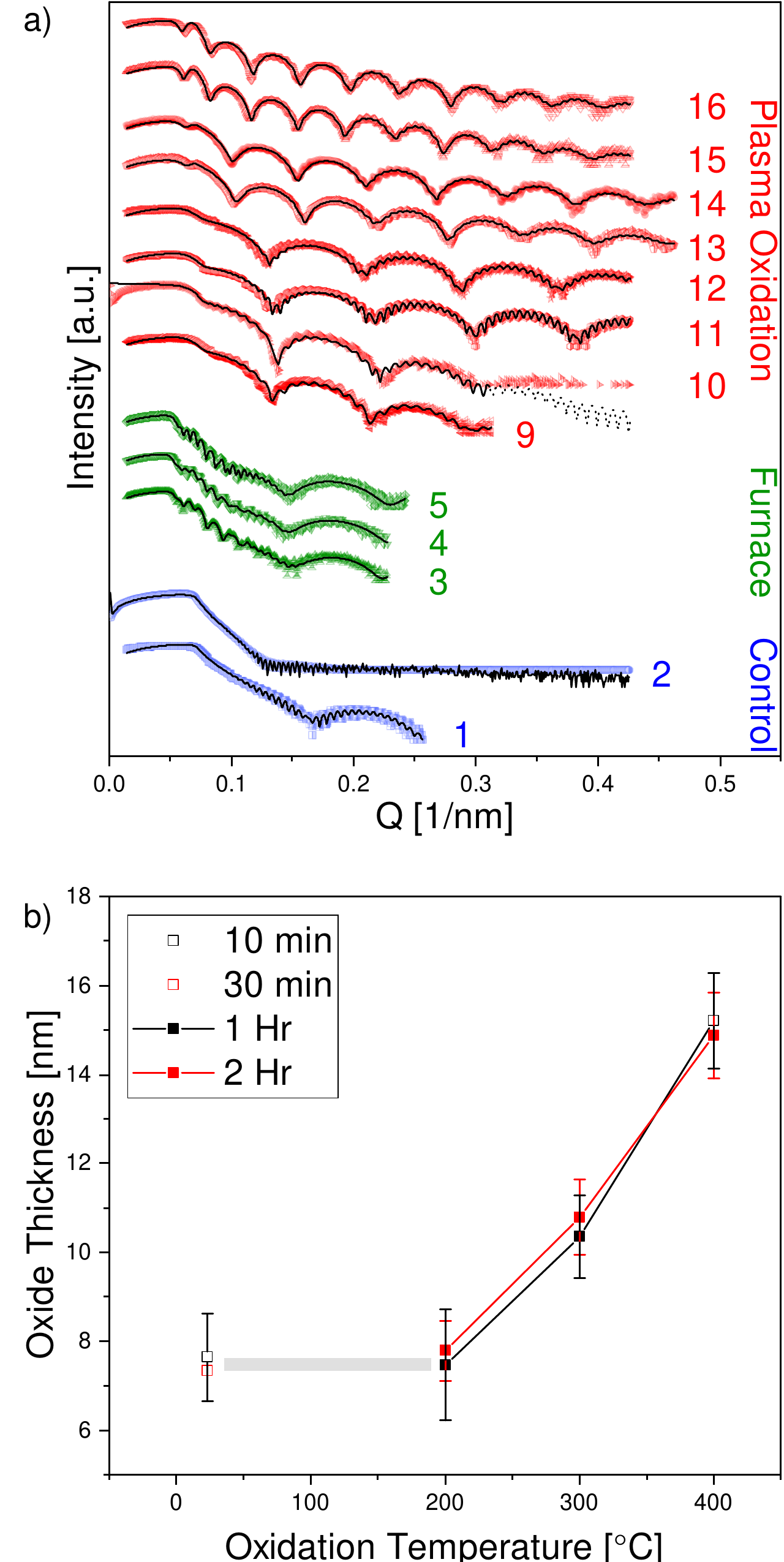}
    \caption{(a) XRR measurements of Ta films subjected to native oxidation (control), tube furnace oxidation, and plasma oxidation treatments. Solid black lines show fitted curves used to extract the oxide thicknesses. The dotted portion of the sample 10 fit curve indicates poor fitting quality at high Q values. (b) Oxidation temperature vs Thickness. A clear trend of approximately equal thickness can be seen for different temperatures and times. The gray line is a guide for the eye.}
    \label{fig:XRR}
\end{figure}

To determine the oxide thicknesses, we employed X-ray reflectometry using a Bruker D8 Discover with a Cu anode X-ray source (50 kV, 1000 \textmu{}A). Fig.~\ref{fig:XRR}(a) displays the measured XRR scattering intensities plotted against $Q$, the momentum transfer vector component perpendicular to the film surface. The solid curves are fits based on the Levenberg-Marquardt model, considering Ta$_2$O$_5$ and Ta layers, as well as the Nb seed layer (when applicable), and appropriate substrate. Note that the films oxidized by rapid thermal annealing exhibited delamination and cracking of the full Ta/Nb/Si stack, as shown in SEM images in Supplementary Information Fig.~S4, and were therefore excluded from XRR. This failure likely originated from residual stress due to the large lattice mismatch between Nb/Ta (both having $a \approx 3.3$ \r{A}) and Si ($a \approx 5.43$ \r{A}),\cite{Moridi2013} combined with the large thermal gradients inherent to RTA. These gradients amplify thermal stresses related to the coefficient of thermal expansion,\cite{Kim2022} leading to structural failure. No such delamination was observed for films oxidized in the tube furnace or via heated plasma.

Based on the XRR fits shown in Fig.~\ref{fig:XRR}(a), the thickness of the oxide was extracted, as summarized in Table \ref{tab:table1}. For our native oxide samples, we extract a thickness of 1.93 and 2.38 nm, which is consistent with previous studies.\cite{McLellan2023, Kim2003}  For the tube furnace annealed films, XRR results show that the thickness of the oxide layer increased from 37 to 60 nm as the annealing time was increased from 10 to 60 min. 

For the plasma oxidized films, we consider the effects of different annealing temperatures as well as annealing times. First, the XRR measurements revealed that exposing the films to oxygen plasma on an unheated stage produced a surface oxide layer of approximately 7–8 nm thick. The extracted thickness for 10 minutes (sample 9) versus 30 minutes (sample 10) of exposure resulted in oxide thicknesses within 0.3 nm, suggesting that the process is insensitive to time within the tested time frames.

To test methods for generating a thicker oxide layer, we further exposed the samples to oxygen plasma while heating them for durations of 1 hour and 2 hours. Plasma oxidation at substrate temperatures of 200$\degree$C, 300$\degree$C, and 400$\degree$C yielded oxide thicknesses of approximately 7–8 nm, 10 nm, and 15 nm, respectively, as extracted from XRR and observed in cross-sectional SEM images in Supplementary Information Fig.~S5. Notably, the extra hour of the oxidation time had negligible influence on the final thickness in each case.

\begin{figure*}[t!]
    \centering
    \includegraphics[width=1\linewidth]{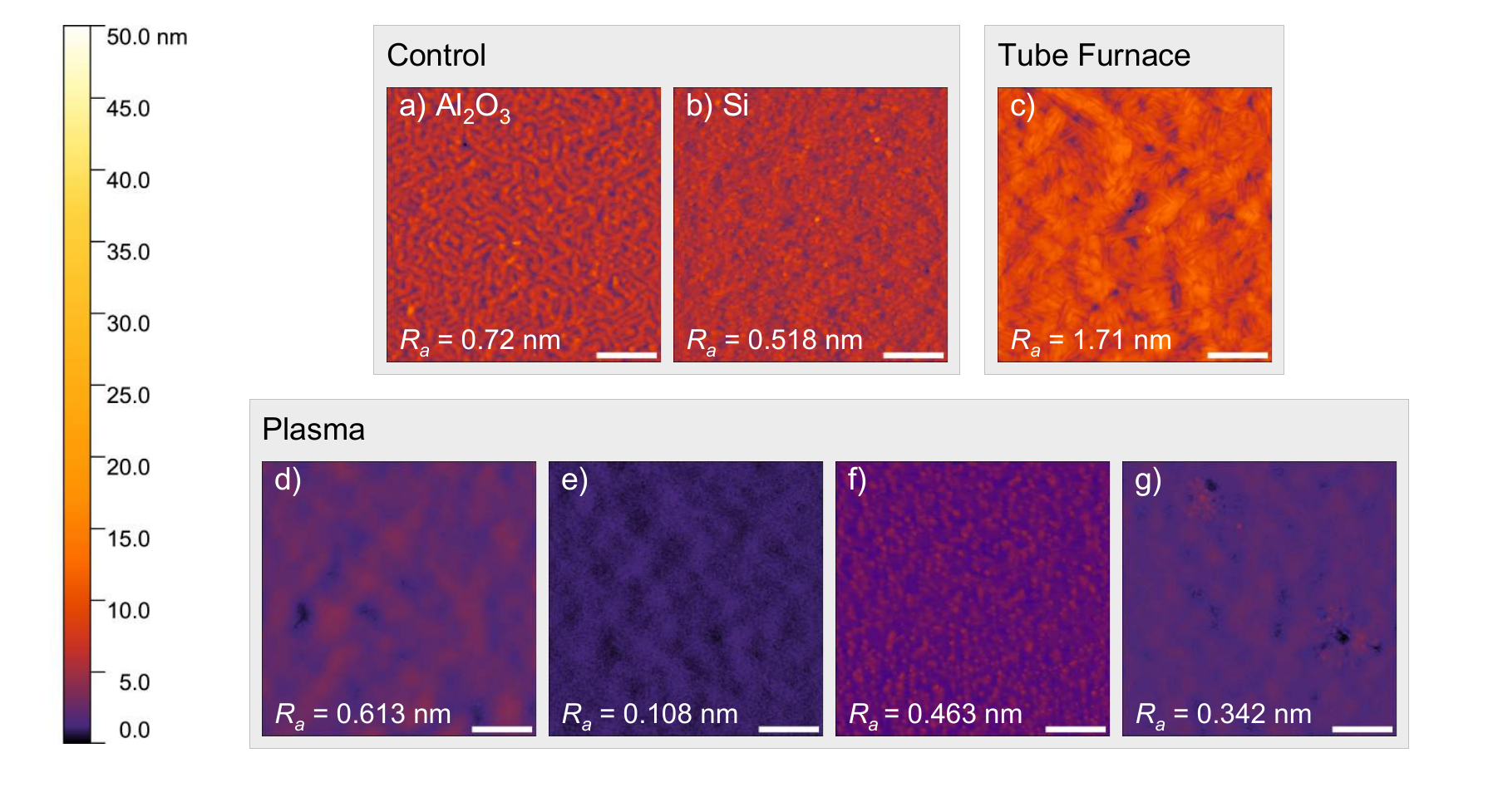}
    \vspace{-1cm}
    \caption{Surface morphology of native and oxidized tantalum films characterized using AFM, with mean roughness ($R_a$) annotated for each sample. Tantalum films grown on (a) sapphire at 500$\degree$C (sample 2) and (b) silicon at room temperature (sample 1) display distinct grain structures. (c) Thermal oxidation (sample 5)
    resulted in an oxide with large grain sizes and moderate surface roughness. Films that underwent plasma oxidation at (d) room temperature (sample 10), (e) 200$\degree$C (sample 11), (f) 300$\degree$C (sample 13), and (g) 400$\degree$C (sample 15) display smooth surfaces with much smaller grains. The color bar shown on the left of the figure corresponds to measured sample height in nm. The scale bars correspond to a length of 200 nm.}
    \label{fig:AFM}
\end{figure*}

Fig.~\ref{fig:XRR}(b) displays the resulting oxide thickness versus annealing temperature. The observed temperature-dependent oxide thickness in our plasma-oxidized samples can be explained by the thermodynamically controlled nature of tantalum oxidation. Ref. [\citenum{Mun2024}] used scanning transmission electron microscopy (STEM), electron energy-loss spectroscopy (EELS), and density functional theory (DFT) calculations to characterize tantalum films oxidized in air, revealing a three-layer structure: amorphous Ta$_2$O$_5$, crystalline TaO$_{1-\delta}$ interface, and unoxidized crystalline Ta. Their analysis showed that oxidation follows a defined pathway in which oxygen penetrates between Ta atomic planes, transitioning from ordered to amorphous structure when the O/Ta ratio exceeds 1:1. The time-independent behavior in our study suggests that each plasma oxidation temperature provides sufficient thermal energy to drive this structural reorganization to completion within 1 hour, with an increase in temperature increasing the effective depth of the O penetration. The final oxide thickness represents the thermodynamic equilibrium state for each temperature, explaining why extending the oxidation time beyond 1 hour does not increase the oxide film thickness -- the system has already reached its most stable configuration.

\subsection{\label{sec:AFM}Surface Morphology}

Superconducting quantum devices with rougher film surfaces may have higher losses than those of smoother films. For example, in one study,\cite{Karuppannan2025} reducing the RMS surface roughness of Nb-based superconducting resonators from 0.98 nm to 0.31 nm quintupled the internal quality factor $Q_i$. This was attributed to a reduction in losses from quasiparticles and TLSs. To quantify the surface roughness of our oxidized Ta films, we performed atomic force microscopy (AFM) scan over $1 \times 1$ \textmu{}m$^2$ and $10 \times 10$ \textmu{}m$^2$ areas. Further details are included in the Methods section. Fig.~\ref{fig:AFM} displays the AFM images from the 1~\textmu{}m$^2$ scans and Table \ref{tab:table1} summarizes the mean roughness extracted from the 10 \textmu{}m$^2$ scans.

The Ta control film grown on a sapphire substrate at high temperature (sample 2) displays three primary domain orientations, evident in Fig.~\ref{fig:AFM}(a). Grains are oriented either in parallel or $\pm60\degree$ relative to each other. Similar grain structures were observed in previous studies of high-temperature epitaxial grown Ta on c-plane sapphire substrates.\cite{Alegria2023, Place2021} In these studies, it was found that the grains aligned with the hexagonal (0001)-oriented c-plane Al$_2$O$_3$ substrate and (110)-oriented Ta crystal. However, the control film grown on silicon with an Nb seed layer (sample 1) displays random grain orientations. Both control films have a low mean surface roughness of 0.518 nm and 0.72 nm, for the silicon and sapphire substrates, respectively. 

\begin{figure*}[t!]
    \includegraphics[width=1\linewidth]{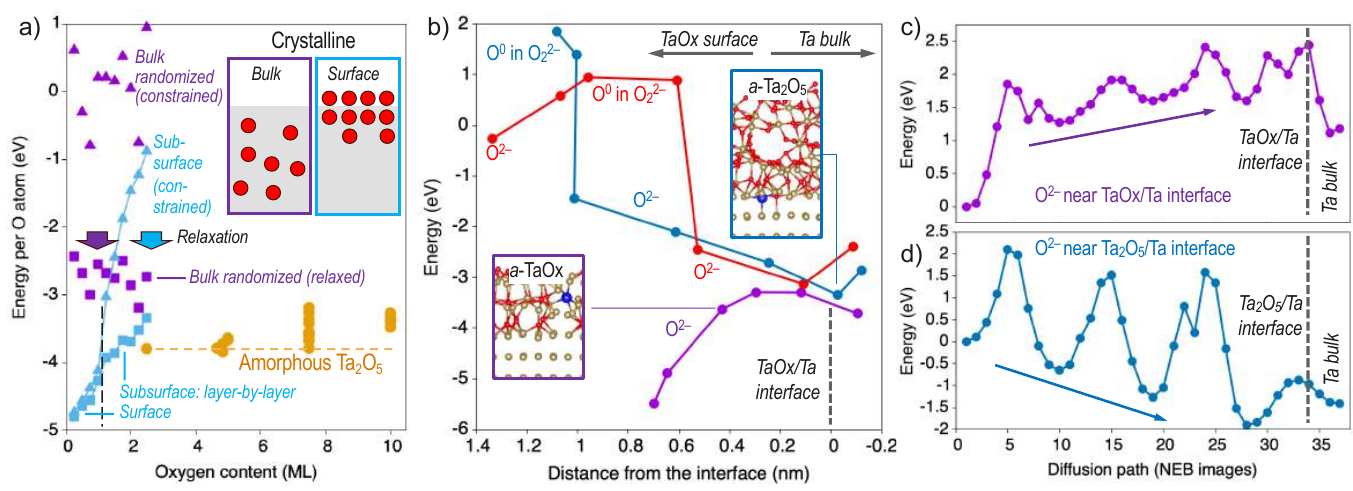}
    \caption{(a) Thermodynamic stability of the TaO$_x$ films supported on the Ta(110) surface depending on the content and spatial distribution of the oxygen species. Isolated O impurities in the bulk Ta are the least stable. Their presence induces lattice relaxation and promotes layer-by-layer accumulation of oxygen, which is further stabilized by the Ta oxidation to Ta$_2$O$_5$ and accompanying amorphization. (b) Energies of oxygen adsorption on the TaO$_x$ surface and absorption into the TaO$_x$ films, depending on the proximity to the interface and film composition. Oxygen adsorbs and diffuses as O$^{2-}$ if the film is not fully oxidized or if the Ta interface is close enough to promote charge transfer, and as O$^0$ bound to a preexisting O$^{2-}$ and forming an O$_2^{2-}$ molecular ion otherwise. (c) and (d) Potential energy profiles for the diffusion of interstitial O$^{2-}$ in the case of partially oxidized TaO$_x$ (c) and nearly fully oxidized amorphous Ta$_2$O$_5$ ($a$-Ta$_2$O$_5$) (d).}\label{fig:OxygenationMechanism}
\end{figure*}

The oxidation procedures examined here resulted in oxide films with distinct surface morphologies and roughness. Thermal oxidation in the tube furnace at 400$\degree$C resulted in the formation of an oxide layer composed of large grains -- approximately 90-130 nm in length and 10-15 nm in width -- that were randomly distributed in clustered formations. Within each cluster, the grains exhibited a nearly parallel alignment. 
The films treated through this process exhibited mean surface roughnesses ($R_a$) of 1.04 to 1.71 nm, indicating a consistently smooth surface morphology.

Plasma oxidation conducted at several temperatures resulted in the smoothest films, with $R_a \approx 0.105-0.613$ nm. Based on the much lower surface roughnesses obtained by this method compared to thermal oxidation in the tube furnace, and ability to conduct the oxidation in-situ, without exposing the films to air, we determined that plasma oxidation is optimal for systematic growth of a smooth, high-quality oxide on tantalum for Josephson junction fabrication.

\subsection{\label{sec:Oxygenation Mechanisms}Ta Oxidation Mechanisms}

To rationalize these results of constant thickness for different temperatures of oxidation, we turn to ab initio modeling and compare the relative stability of partially oxidized Ta depending on the concentration and spatial distribution of the oxygen species. We considered three distinct scenarios for the oxygen species: (1) interstitial O injected into Ta disperse throughout the bulk; (2) oxygens sequentially occupy surface binding sites and subsurface interstitial sites at a 1:1 ratio with Ta, thus forming a TaO composition in the near-surface region; (3) oxygens are randomly distributed over the volume occupied by the surface Ta planes so as their ratio Ta:O=2:5. In the latter case, the lattice relaxations results in amorphous structures. Comparison of the calculated energy gain per O atom vs oxygen content for these scenarios [see Fig.~\ref{fig:OxygenationMechanism}(a)] shows that fully oxidized amorphous Ta$_2$O$_5$ is more stable than Ta with a spread-out oxygen distribution or quasi-ordered layer-by-layer TiO structures. Thus, isolated interstitial oxygen atoms are unlikely to diffuse from the surface into the Ta bulk. However, these atoms can be readily incorporated via oxygen plasma treatment, which would skew the native oxide formation. 

For the plasma process, oxygenation of tantalum begins with the injection of oxygen ions, through the tantalum native oxide layer, penetrating to a depth defined by the stopping range --- an extent that depends directly on the plasma power used during the process. Once implanted, these ions undergo thermal diffusion -- our calculations suggest that the barrier for this diffusion is $\sim$1 eV [see the last barrier in Fig.~\ref{fig:OxygenationMechanism}(d)] --  and spread out over the Ta lattice. However, as their concentration increases, the strain buildup associated with interstitial atoms relaxes via the lattice expansion in the out-of-plane direction. This relaxation increases spacing between the Ta planes and promotes a layer-by-layer oxygen accumulation and eventual formation of amorphous Ta$_2$O$_5$ ($a$-Ta$_2$O$_5$) [see Fig.~\ref{fig:OxygenationMechanism}(a)].

We note that oxidation due to plasma treatment is accompanied by the oxygen intake from the surface. At low oxygen content, the adsorption and absorption of additional O species is thermodynamically preferred (purple line in  Fig.~\ref{fig:OxygenationMechanism}(b)) and their diffusion is facile with the calculated barriers of $\sim$0.5 eV [Fig.~\ref{fig:OxygenationMechanism}(c)]. The uphill slope in Fig.~\ref{fig:OxygenationMechanism}(c) indicates the thermodynamic preference to saturate oxygen content locally, i.e., form Ta$_2$O$_5$, before diffusing toward the Ta metal.

As the Ta oxidizes to TaO$_x$, the oxygen deposition location changes due to the differences in the Ta and TaO$_x$ stopping ranges and the film expansion; eventually, the newly injected oxygen ions are deposited within the oxide itself. If these oxygens are deposited in the vicinity of the interface between a nearly fully oxidized Ta$_2$O$_5$ and Ta metal, they draw the electron charge from the Ta substrate, convert to O$^{2-}$ [Fig.~\ref{fig:OxygenationMechanism} (b)], and diffuse toward the interface driven by the positive charge associated with the interface. This trend is illustrated by the negative slope in (Fig.~\ref{fig:OxygenationMechanism} (d)). Importantly, the slope changes sign at the interface, indicating that further O$^{2-}$ diffusion into metal Ta is suppressed\cite{Mun2024} even though the barrier for diffusion decreases from $\sim$2 eV to $\sim$1 eV.

As the lattice expansion proceeds and the local Ta:O ratio reaches 2:5, the new oxygen species are no longer deposited near the Ta interface. In this case, they form peroxy species (see Supplementary Information section S4), i.e., a neutral atom O$^0$ binding to the lattice O$^{2-}$ and forming an O$_2^{2-}$ molecular ion. (For ball-and-stick representations of the local atomic configurations of excess oxygen in Ta$_2$O$_5$ and associated energy profile, see Supplementary Information Fig.~S6). At low concentrations, these neutral oxygen species do not have a preferential diffusion direction. Since their formation is unfavorable relative to the O$_2$ gas-phase molecule [blue and red lines in Fig.~\ref{fig:OxygenationMechanism}(b)], they are likely to escape the Ta$_2$O$_5$ oxide layer through out-diffusion.

Consequently, the final thickness of the oxide is governed by the plasma power, which sets the initial penetration depth and saturation time that depend on the kinetic energy of the plasma species and flux, and the temperature, which controls the extent of subsequent thermal diffusion.

As the process continues, any additional oxygen may either remain interstitial or diffuse out of the film.
To compare the plasma and tube furnace oxygenation processes, note that the tube furnace operated at an oxygen flow rate nearly 1000 times higher and produced significantly thicker oxide layers. This elevated flow rate likely contributes to a high oxygen chemical potential, which promotes the formation of interstitial oxygen atoms and their diffusion deeper into the film owing to the high concentration gradients. For more details, refer to the discussion on the interstitial O$^0$ atom diffusion in $a$-Ta$_2$O$_5$ in Supplementary Information section S4.

\subsection{\label{sec:Trilayers}Trilayers}

To demonstrate the potential for Ta-based JJs, we test procedures to grow $\alpha$-Ta on top of this Ta$_2$O$_5$ layer, forming a SIS trilayer stack suitable for junction fabrication. An earlier effort to grow similar trilayers through heated deposition of Ta, reported in Ref. [\citenum{Lam2006}], found that the top layer above the oxide contained polycrystalline $\alpha$- and $\beta$-phase Ta. The oxide was also suspected to contain pinholes that produced microshorts between the top and bottom Ta layers. Recognizing from our Ta film growth and characterization that $\alpha$-Ta can be achieved either via nucleated growth using a Nb seed layer at room temperature or by direct deposition at 500$\degree$C without a seed layer, we evaluated both approaches to determine their effectiveness in producing 
$\alpha$-Ta on our oxide surfaces.

For one process, we sputtered a 6-nm Nb seed layer, followed by 100 nm of Ta on top of cuts of a Ta film with a native oxide (sample 1) and two Ta films that were oxidized in an oxygen plasma (samples 15 and 16). For the second process, we sputtered 150 nm of Ta on top of cuts of samples 1, 2, 15, and 16 held at 500$\degree$C. The deposition rates, argon process pressure, and DC power are identical to those reported in the Methods section for the initial $\alpha$-phase Ta deposition process.

\subsubsection{X-ray Diffraction on Trilayers}

\begin{figure}[t!]
    \centering
    \includegraphics[width=1\linewidth]{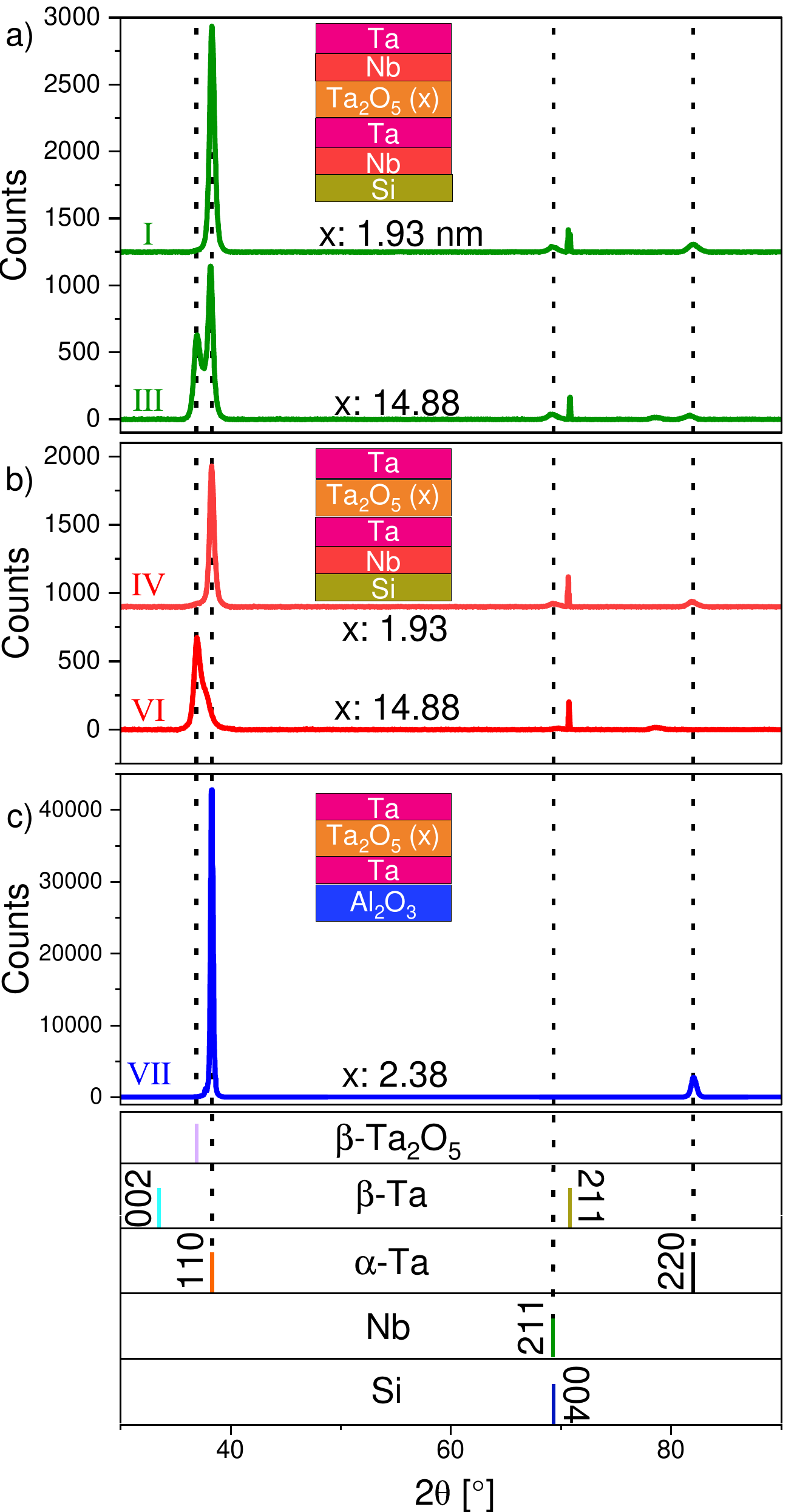}
    \caption{(a-c) X-ray diffraction (XRD) patterns of the trilayer films. The stack order for each set is shown in the cartoon insets, with "x" indicating the thickness of the oxide in nm. (Bottom panel) Reference peak angles for the phases of interest:  $\beta$-Ta$_2$O$_5$ ($36.9\degree$),\cite{Bright2013}  $\alpha$-Ta (110, $38.3\degree$),\cite{Wu2023} and (220, $82\degree$),\cite{Wu2023} Nb (211,  $69.3\degree$),\cite{Eusterholz2023, Li2016} and Si (004,$69.3\degree$).\cite{Bing-Hwai2001} as measured in our films.}
    \label{fig:XRD}
\end{figure}

Fig.~\ref{fig:XRD} shows the XRD spectra for these samples. Trilayers labeled as I and III were grown at room temperature with an Nb seed layer on samples 1 (control) and 16 (plasma oxidized), respectively. For the trilayers labeled IV and VI, the top Ta layer was deposited at 500$\degree$C without a seed layer. Lastly,

\begin{figure*}[t!]
    \includegraphics[width=1\linewidth]{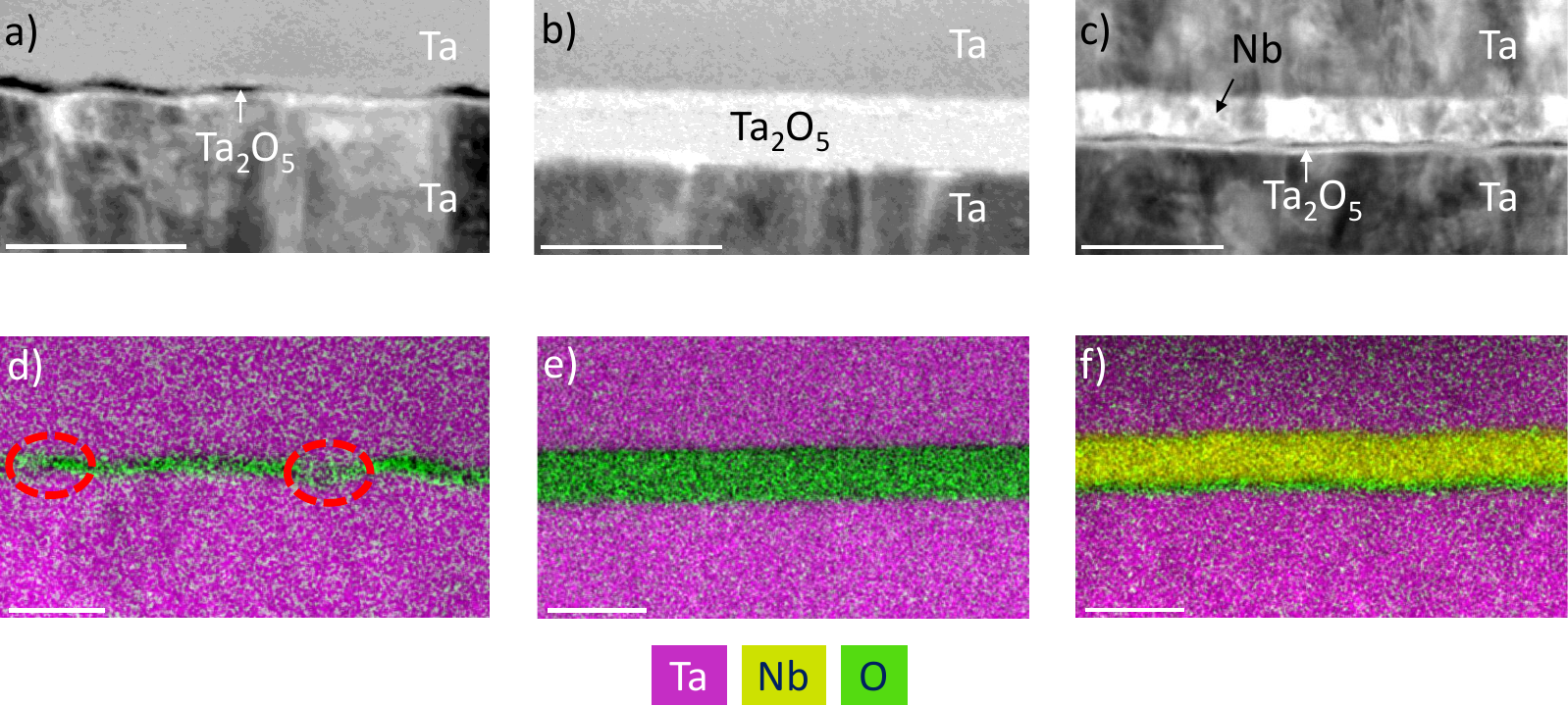}
    \caption{[Top row] STEM-HAADF overview images of three different trilayer samples, corresponding to (a) sample \RomanNumeral{7} and (b) \RomanNumeral{5}, as labeled in Fig.~\ref{fig:XRD}, as well as (c) sample \RomanNumeral{1} (XRD absent from Fig.~\ref{fig:XRD}). [Bottom row, d - f] STEM-EDX elemental mapping of Ta, O, and Nb (for sample \RomanNumeral{1}). Scale bars correspond to a length of 50 nm.}
    \label{fig:STEM}
\end{figure*}

\noindent VII labels the trilayer in which Ta was also grown at 500$\degree$C without a seed layer, but on sample 2 (control on sapphire). The trilayer labeling scheme is summarized in Table \ref{tab:trilayers}.

\begin{table}
\caption{\label{tab:trilayers} Trilayer samples including the sample ID, layers, substrate, and tantalum-oxide thickness.}
\begin{ruledtabular}
\begin{tabular}{lccc}
 ID & Stack & Substrate & TaO$_x$ thickness [nm]  \\
\hline
I &  &  & 1.93\\
II & Ta/Nb/Ta$_2$O$_5$/Ta/Nb & Si & 15.21\\
III &  &  & 14.88\\
\hline
IV &  &  & 1.93\\
V & Ta/Ta$_2$O$_5$/Ta/Nb & Si & 15.21\\
VI &  &  & 14.88\\
\hline
VII & Ta/Ta$_2$O$_5$/Ta & Al$_2$O$_3$ & 2.38\\
\end{tabular}
\end{ruledtabular}
\end{table}

Focusing first on the trilayers shown in panel (a), where tantalum growth on Ta$_2$O$_5$ was facilitated by an Nb seed layer (green curves labeled I and III), we observe distinct reflections corresponding to the (110) and (220) planes of $\alpha$-Ta. In the sample with a 14.88 nm oxide layer, a pronounced peak attributed to $\beta$-Ta$_2$O$_5$ emerges at a 2$\theta$ angle of approximately 36.9$\degree$, while in the sample with a thinner 1.93 nm oxide, the magnitude of this peak is significantly reduced, manifesting only as a broadening of the adjacent (110) $\alpha$-Ta peak. The typical prominent (002) peak of $\beta$-Ta that appears in pure $\beta$-Ta\cite{Palmstrom2025} is notably absent, suggesting that we have negligible $\beta$-phase Ta in either layers. Regarding potential contributions from the Nb seed layer, it is important to note that the (211) reflection of Nb closely aligns with the (004) Si substrate peak, resulting in noticeable broadening around $69\degree$-$70\degree$ in the spectra. The sharp peak at $70.7\degree$ is attributed to defects in the substrate, which can also be seen in the XRD measurement of our Si substrates, shown in Supplementary Information Fig.~S1(b).

 Fig.~\ref{fig:XRD}(b) displays the X-ray spectra for films on Si substrates in which tantalum was deposited at 500$\degree$C directly onto the Ta$_2$O$_5$ layer, without an Nb seed layer. While the same crystalline phases appear as in the previously discussed stack with Nb atop the oxide, the relative intensities of the peaks differ. Notably, the (110) $\alpha$-Ta peak is significantly reduced, appearing as a broad shoulder for the $\beta$-Ta$_2$O$_5$ peak. 

Lastly, Fig.~\ref{fig:XRD}(c) presents data for a trilayer in which the top tantalum layer was deposited directly onto Ta$_2$O$_5$ at 500$\degree$C. Unlike the samples in panel (b), the bottom Ta layer in this structure was grown directly on a sapphire substrate, rather than on Nb/Si. The X-ray spectra reveal only the (110) and (220) reflections characteristic of $\alpha$-Ta. 

To determine whether the $\alpha$-Ta peaks reflect the crystalline phase of both Ta layers or only the bottom layer, we measured the sheet resistance of the top layer using a four‑point probe. From these measurements, we calculated the electrical resistivity of the top Ta layer, assuming negligible conduction through the insulating Ta$_2$O$_5$. For trilayer III, which has an Nb seed layer below both Ta layers, we measured a resistivity of 23.7392 \textmu{}$\Omega$-cm. This value is within the typical range of resistivities for $\alpha$-Ta (10-60 \textmu{}$\Omega$-cm).\cite{Stella2009,Zhang2003,Valleti2008} We show additional evidence that both Ta layers in trilayer III are of the $\alpha$ phase through DC magnetization measurements, which show a critical temperature of 4.5 K and an approximate doubling of the magnetic moment in the superconducting state, as compared to a single-layer Ta film grown with an Nb seed layer on a Si substrate (See Supplementary Information Fig. S1). For a sample with the top Ta layer grown directly on the Ta$_2$O$_5$ (trilayer VI), we measured a resistivity of 191.1 \textmu{}$\Omega$-cm. This value falls within the previously measured range of resistivities for $\beta$-Ta (100-200 \textmu$\Omega$-cm).\cite{Stella2009,Feinstein1972,Nnolim2003,Schwartz1972,Sajovec1992} and amorphous Ta (>200 \textmu{}$\Omega$-cm)\cite{Stella2009,Narayan2006}. Given the lack of a $\beta$-Ta (002) peak in the XRD, the top layer is likely largely amorphous, further investigated in the electron microscopy studies presented in Section \ref{sec:TEM}.

\begin{figure*}
    \includegraphics[width=1\linewidth]{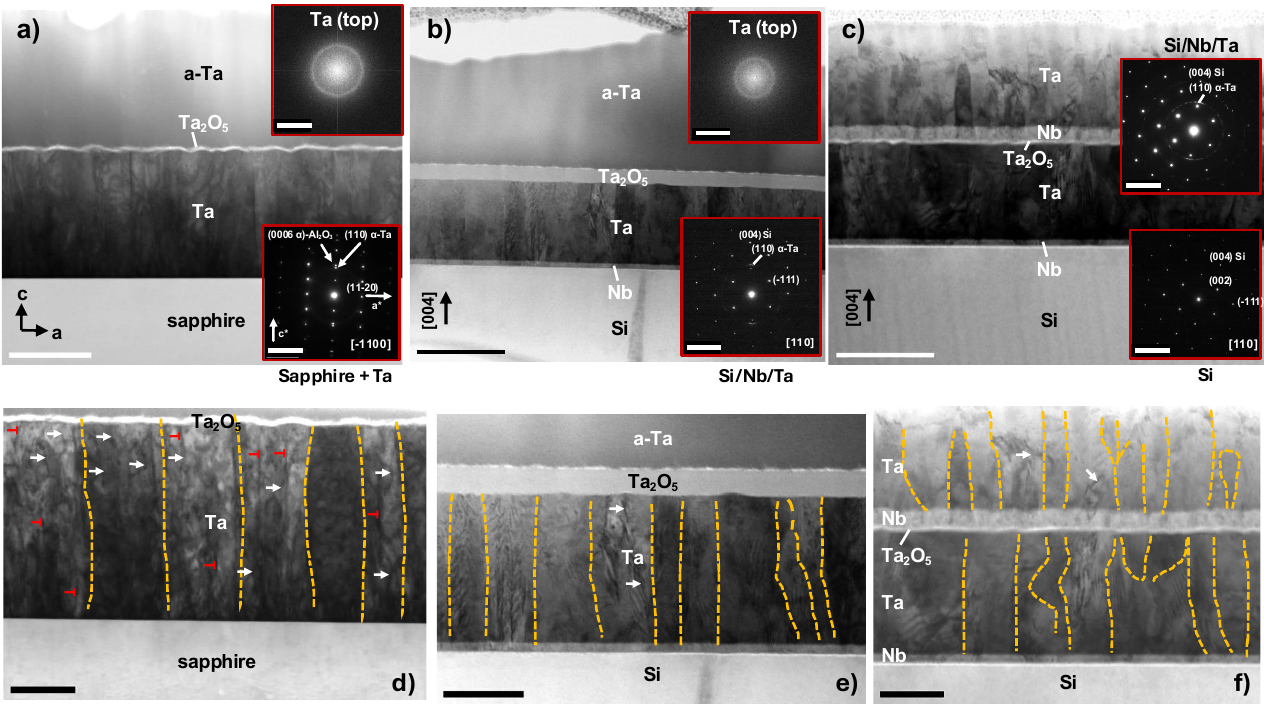}
    \caption{Cross-sectional STEM-BF images of three different trilayer samples: (a) Ta/Ta$_2$O$_5$/Ta (sample \RomanNumeral{7}) on sapphire, (b) Ta/Ta$_2$O$_5$/Ta on Nb/Si (sample \RomanNumeral{5}), and (c) Ta/Nb/Ta$_2$O$_5$/Ta on Nb/Si (sample \RomanNumeral{1}). Insets show the FFT and SAED patterns from the indicated layers, confirming the crystallinity and orientation of the substrate and trilayers. For the SAED patterns, the zone axes is noted in the lower left corner. The scale bars are 100 nm and 5 nm$^{-1}$ for the STEM images and SAED/FFT, respectively. (d–f) High-magnification STEM-BF images highlighting columnar grains and grain boundaries (yellow dashed lines), threading dislocations (red arrows), and mixed dislocations (white arrows). The scale bars correspond to 50 nm.
}
    \label{fig:HighResSTEM}
\end{figure*}

\subsubsection{\label{sec:TEM}Transmission Electron Microscopy of Trilayers}

To evaluate the interface quality, we imaged lamellae, prepared by focused ion beam-scanning electron microscopy (FIB-SEM), of three trilayer samples using scanning transmission electron microscopy (STEM), including high-angle annular dark-field (HAADF) imaging and energy-dispersive X-ray (EDX) spectroscopy, shown in Fig.~\ref{fig:STEM}. More details regarding the electron microscopy work are provided in the Methods section.

STEM‑HAADF imaging of trilayer \RomanNumeral{7}, in which Ta is deposited atop a native oxide, shows two Ta layers --- 168 nm (top) and 160 nm (bottom) --- encapsulating a Ta$_2$O$_5$  layer with a non‑uniform thickness of 4–8 nm and associated interfacial roughness, with a magnified view shown in Fig.~\ref{fig:STEM}(a). Corresponding EDX elemental mapping in Fig.~\ref{fig:STEM}(d) shows small Ta-rich regions within the barrier (circled in red), which may behave as superconducting shorts between the top and bottom Ta layers.

Fig.~\ref{fig:STEM}(b, e) display the STEM-HAADF and EDX images of trilayer \RomanNumeral{5}, in which Ta was grown on a thicker oxide layer. The Ta$_2$O$_5$ layer appears uniform, 15 nm thick, and has sharp interfaces. Lastly, Fig.~\ref{fig:STEM}(c) presents the STEM-HAADF image for the only trilayer in which the top layer was grown at room temperature and on a Nb seed layer -- sample \RomanNumeral{1}. EDX mapping in part (f) reveals a 2–4 nm-thick Ta$_2$O$_5$ layer situated between two Ta layers, each $\sim$ 100-105 nm thick. The Nb layer used to nucleate the $\alpha$-phase top electrode measures 13 nm (unintentionally thick), whereas the Nb seed layer on the substrate is 5-6 nm thick.

For all samples, EDX analysis detected oxygen signals throughout the top and bottom Ta layers (see Supplementary Information Fig.~S7). This contamination of oxygen in the Ta bulk likely occurred during exposure of the FIB-cut lamella to air prior to STEM imaging and EDX mapping. Consequently, we are unable to evaluate the propensity for oxygen incorporation from any residual chamber oxygen contamination.

To characterize the defect landscape and evaluate the crystallinity of the trilayers, the lamellae were further thinned and examined using higher‑resolution S/TEM microscopy. Fig.~\ref{fig:HighResSTEM}(a–c) provides a comparative structural analysis of trilayer samples grown on sapphire (0001) and silicon (004) substrates. Cross‑sectional STEM bright‑field (STEM‑BF) imaging, together with selected‑area electron diffraction (SAED) and fast Fourier transform (FFT) analysis (insets), was used to assess crystallinity and orientation relationships between the substrates and trilayers for three samples: Ta/Ta$_2$O$_5$/Ta on sapphire (trilayer~\RomanNumeral{7}), Ta/Ta$_2$O$_5$/Ta on Nb/Si (trilayer~\RomanNumeral{5}), and Ta/Nb/Ta$_2$O$_5$/Ta on Nb/Si (sample~\RomanNumeral{1}).

For trilayer~\RomanNumeral{7}, the SAED pattern acquired at the sapphire/Ta interface confirms that the sapphire c‑axis is oriented normal to the substrate surface. The diffraction patterns further reveal a well‑defined epitaxial relationship between the sapphire substrate and the Ta layer. Specifically, SAED indicates the orientation relationship (110)$\alpha$‑Ta ‖ (0001) Al$_2$O$_3$, with columnar grains extending along the sapphire c‑axis throughout the Ta layer. This textured growth of the Ta layer on sapphire, following this orientation relationship, results in pronounced columnar grains.    Regarding the top Ta layer in trilayer \RomanNumeral{7}, it was notably found to be amorphous, as evidenced by the FFT analysis, shown in the top inset of Fig.~\ref{fig:HighResSTEM}(a).

For trilayer \RomanNumeral{5}, the SAED pattern acquired at the Si substrate/Ta interface (Fig.~\ref{fig:HighResSTEM}(b), bottom inset) shows the Ta (110) reflection as arcs, consistent with textured growth primarily along the Si [004] direction and indicative of preferential columnar Ta growth. In contrast to trilayer \RomanNumeral{7}, however, the image also reveals a population of misoriented grains. FFT analysis of the top Ta layer (Fig.~\ref{fig:HighResSTEM}(b), top inset) further confirms that this layer is amorphous.

The final trilayer imaged (trilayer \RomanNumeral{1}) incorporates an Nb layer above the oxide. For this sample, the lower inset of Fig.~\ref{fig:HighResSTEM}(c) displays the SAED pattern of the Si substrate acquired along the [110] zone axis, with the (004) reflection oriented normal to the surface. The SAED patterns from the Si and Ta layers exhibit single‑crystal spots from Si and diffraction rings from Ta layers, indicating grains with multiple orientations and confirming the polycrystalline character of the Ta layer. These observations show that introducing an Nb interlayer above the Ta$_2$O$_5$ barrier promotes crystallization and grain growth in the top Ta layer.

To identify the types of defects present, high‑magnification STEM‑BF images are shown in Fig.~\ref{fig:HighResSTEM}(d–e). These images reveal columnar grains separated by well‑defined grain boundaries (yellow dashed lines) and a high density of threading (red arrows) and mixed dislocations (white arrows) in trilayer~\RomanNumeral{7}  (Fig.~\ref{fig:HighResSTEM}(d)). The dislocation density in trilayer \RomanNumeral{7} is substantially higher than in trilayers \RomanNumeral{5} and \RomanNumeral{1}. In contrast, trilayers \RomanNumeral{5} and \RomanNumeral{1} display more pronounced Moir\'e fringes, indicative of increased grain misorientation. Additional electron microscopy results—including dark‑field, HRTEM, and bright‑field images—are provided in Supplementary Information Section S6 (Figs. S8–S10).

In summary, all trilayers exhibit c‑axis‑oriented columnar grain growth of the bottom Ta layer. Films grown on sapphire substrates show larger grains, fewer misoriented grains, and a higher density of dislocations than those grown on silicon. Nucleating c-axis-oriented $\alpha$-Ta on Ta$_2$O$_5$ required an Nb seed layer—direct deposition of Ta on the oxide yielded amorphous Ta. 

\section{\label{sec:Conclusions}Conclusions}

Through thermal oxidation in a tube furnace, rapid thermal annealing, and plasma oxidation, we grew Ta$_2$O$_5$ on top of sputtered $\alpha$-Ta thin films. Among these approaches, plasma oxidation yielded especially smooth oxide surfaces, with thickness controllable through the annealing temperature. The mechanistic insights into the Ta oxidation process provide guidance for the modulation of the Ta oxide composition and thickness by on-the-fly tuning the plasma parameters. We further demonstrated successful growth of $\alpha$-Ta on Nb-seeded tantalum-oxide atop $\alpha$-Ta films,  establishing a route toward trilayer Josephson junctions incorporating Ta$_2$O$_5$ barriers.

Our results show that Ta/Ta$_2$O$_5$/Ta trilayers can be grown entirely in situ, without breaking vacuum, provided that an Nb seed layer is introduced above the oxide barrier.  These structures provide a promising platform for fabricating highly oriented Ta‑based Josephson junctions, enabling quantitative studies of TLS defect densities in the barrier and interfaces and allowing integration into qubits to compare coherence and lifetimes with state‑of‑the‑art Ta devices employing conventional Al‑based junctions.

\section{\label{sec:methods} Methods\protect}

\subsection{DC Magnetron Sputtering of Tantalum Films\protect}

We employed two distinct methods to deposit $\alpha$-phase tantalum films. For both procedures, the chamber base pressure reached below $6\times10^{-8}$ torr prior to heating. All substrates were pre-cleaned using acetone and isopropanol, then dried with a nitrogen (N$_2$) spray.

In one approach, we thermally degassed 430 \micro m-thick, double-side polished sapphire substrates pre-heated to $500\degree$C (at a rate of $20\degree$C per minute) for 2 hours. Following degassing, we deposited a 150 nm Ta layer while maintaining the substrate temperature at $500\degree$C. The substrates were then cooled to room temperature at a rate of approximately $12\degree$C per minute.

The second method involved sputtering tantalum at room temperature onto Si(100) substrates with thicknesses of 500 - 695 \micro m. We deposited a 6 nm niobium seed layer, followed by 60 or 100 nm of tantalum. The DC sputtering powers were set to 510 W for Nb and 310 W for Ta. Both materials were deposited under an argon process pressure of 3 mtorr, yielding a deposition rate of approximately 15 nm per minute. 

\subsection{X-ray Spectroscopy, Diffraction, and Reflectometry\protect}

We acquired XPS spectra using a Kratos Axis-Ultra DLD spectrometer equipped with a monochromatized Cu K$\alpha$ X-ray source and a low-energy electron flood gun for charge neutralization. During spectral acquisition, the analytical chamber pressure remained below 
$5 \times 10^{-9}$ torr. Survey and compositional spectra were recorded at a pass energy of 80 eV, while high-resolution spectra were collected at a pass energy of 20 eV. The take-off angle—defined as the angle between the sample normal and the energy analyzer's input axis—was set to 0\degree, corresponding to an approximate sampling depth of 100 \AA.
The Kratos Vision2 software program was used to determine peak areas and to calculate the elemental compositions from peak areas. CasaXPS was used to peak fit the high-resolution spectra. For the high-resolution spectra, a Shirley background was used, and all binding energies were referenced to the C ls C-C bonds at 285.0 eV.

XRD analysis was done using Bruker D8 Discover with a Cu anode
X-ray source (50 kV, 1000 \textmu{}A), combined with a Pilatus 100K 2D detector. The experimental XRD data was then processed using DIFFRAC.EVA software package.

For XRR analysis, the same instrument with the same X-ray source, equipped with a Lynxeye-XET multi-strip 1D detector was used, with a 0.2 mm beam collimator. The experimental XRR data were fitted to a theoretical model using the DIFFRAC.XRR software package. The layer thicknesses, densities, and interface roughnesses were varied to achieve the best fit to the experimental data. The goodness of fit was determined by minimizing the chi-square \(\chi ^{2}\) value between the measured and simulated reflectivity curves.

\subsection{Atomic Force Microscopy (AFM)\protect}
Atomic force microscopy was performed using a Bruker ICON AFM, with a Bruker ScanAsyst-Air probe with a 2 nm tip in ScanAsyst mode and a Bruker OTESPA-R3 probe with a 7 nm tip in tapping mode. The 1 \textmu{}m$^2$ scans were collected at a rate of 1 Hz with 512 samples per line, and the 10 \textmu{}m$^2$ scans were collected at a rate of 1 Hz with either 256 or 512 samples per line.

\subsection{Electron Microscopy}

TEM samples were prepared using an FEI Helios NanoLab 600i Focused Ion Beam Scanning Electron Microscope (FIB/SEM). Layer stacking and elemental composition were examined with a Talos F200X STEM operated at 200 kV, equipped with a field emission gun (FEG) and four in‑column Super‑X EDS detectors. STEM-HAADF (high angle annular dark-field ) images were recorded using a convergence semi angle of 10.5 mrad.

\subsection{Computational Modeling}

The dependence of the thermodynamic stability of Ta oxide films supported on the Ta substrate on the content and spatial distribution of the oxygen species was quantified using ab initio simulations within the density functional theory (DFT) formalism, and a generalized gradient approximation exchange correlation functional\cite{Perdew1996} as implemented in the VASP code.\cite{Kresse1996} The plane wave augmented pseudopotentials were used.\cite{Bloch1994} The Ta slab was simulated using the periodic model approach. The slab was terminated with (110) surfaces and included 8 atomic planes; the lateral supercell (9.33$\times$9.90 \AA$^2$) included 12 Ta atoms in each plane. The out-of-plane supercell parameter was fixed at 42 \AA, which leaves at least 15 \AA\ wide vacuum gap even after accounting for the slab expansion due to Ta oxidation. The structural models of the amorphous Ta$_2$O$_5$ ($a$-Ta$_2$O$_5$) and Ta oxide (TaO$_x$) film on the Ta surface were generated by incorporating oxygen atoms into the Ta slab. 

Since Ta$_2$O$_5$ and TaO$_x$ are disordered, the simulations were conducted at the $\Gamma$ point only; the plane wave basis set was fixed at 500 eV. The total energies were minimized with respect to the internal atomic coordinates, except for the Ta atoms farthest from the interface plane, which remained fixed at their crystallographic sites. The total energy conversion criterion was set to 10$^{-5}$ eV. Energies of oxygen interaction with Ta and Ta oxides are calculated relative to the energy of the gas-phase O$_2$ molecule. The energy barriers for oxygen diffusion were calculated using the nudged elastic band (NEB) method.\cite{Henkelman2000} The atomic charges were analyzed using the method developed by Bader.\cite{Tang2009_Bader} 

\section*{Data Availability Statement}

The data supporting the findings of this study are available on Mendeley Data (NOT PUBLISHED FOR arXiv VERSION) as a zip file. This includes Origin files (.opju) that contain data spreadsheets for all the samples and figures used in this paper, which can be opened using Origin Viewer, a free application that permits viewing and copying of data contained in Origin project files.

\section*{References}

\bibliography{aipsamp}

\begin{acknowledgments}
We thank Dr. Samantha Young, a staff scientist at the University of Washington's Molecular Analysis Facility (MAF), for collecting the XPS spectra, and for many insightful discussions about X-ray analyses. 
\end{acknowledgments}

\section*{Funding}
This work was primarily supported by the U.S. Department of Energy (DOE), Office of Science, National Quantum Information Science Research Centers, Co-design Center for Quantum Advantage (C2QA) under Contract No. DE-SC0012704 and PNNL FWP 76274 (R.P., R.T., P.V.S., S.E.). Materials growth and processing were conducted at the Washington Nanofabrication Facility / Molecular Analysis Facility, a National Nanotechnology Coordinated Infrastructure (NNCI) site at the University of Washington with partial support from the National Science Foundation via awards NNCI-1542101 and NNCI-2025489. Focused Ion Beam-Scanning Electron Microscopy (FIB-SEM) and Scanning Transmission Electron Microscopy with Energy Dispersive X-ray Spectroscopy (STEM-EDX) were performed in the following core facility, which is part of the Colorado School of Mines Shared Instrumentation Facility: Electron Microscopy (RRID:SCR\_022048). This research also used resources of the National Energy Research Scientific Computing Center (NERSC), a DOE Office of Science User Facility supported by the Office of Science of the U.S. DOE under Contract No. DE-AC02-05CH11231 using NERSC award BES-ERCAP0033525. This work was authored in part by the National Laboratory of the Rockies for the DOE, operated under Contract No. DE-AC36-08GO28308. The views expressed in the article do not necessarily represent the views of the DOE or the U.S. Government.  

\section*{Author Contributions}

S.E. and D.P.P. conceived and designed the experiment.
R.P. and R.T. grew the tantalum films and performed the oxidations, X-ray reflectometry, atomic force microscopy, scanning electron microscopy, and magnetometry measurements.
S.B. collected some x-ray reflectometry data and provided advice on extracting accurate oxide-layer thicknesses. 
TEM specimens (lamellae) were prepared using the FIB lift‑out in cross‑section by A.B. P. K. performed the STEM‑EDX measurements and carried out the data interpretation of these samples.
J.L. conducted cross-sectional SEM imaging, optical microscopy, and magnetometry.
R.P. performed data analysis.
P.V.S. determined the oxygenation pathways and oxidation mechanisms of the Ta films. 
S.E., R.P., R.T., and P.V.S. wrote the manuscript.
All authors commented on the manuscript.

\section*{Competing interests}
The authors have no conflicts to disclose.

\end{document}


\setcounter{figure}{0}
\makeatletter 
\renewcommand{\thefigure}{S\@arabic\c@figure}
\makeatother

\vspace*{-2em}
\begin{center}
    {\Large \textbf{Supplementary Information:}}\\[0.4em]
    {\Large \textbf{Fabrication and Structural Analysis of Trilayers for Tantalum Josephson Junctions with Ta$_2$O$_5$ Barriers}}\\[1.0em]
    Raahul Potluri$^{1}$, Rohin Tangirala$^{1}$, Jiangteng Liu$^{1}$, Alejandro Barrios$^{2}$, Praveen Kumar$^{2}$, Sage Bauers$^{3}$, Peter V. Sushko$^{4}$, David P. Pappas$^{5}$, and Serena Eley$^{1}$
    \\[0.8em]
    {\small
    $^{1}$Department of Electrical \& Computer Engineering, University of Washington, Seattle, WA 98195, USA\\
    $^{2}$Shared Instrumentation Facility, Colorado School of Mines, Golden, CO 80401, USA\\
    $^{3}$National Laboratory of the Rockies, Golden, CO, USA 80401\\
    $^{4}$Physical and Computational Sciences Directorate, Pacific Northwest National Laboratory, Richland, WA 99354, USA\\
    $^{5}$Rigetti Computing, Inc., Berkeley, CA 94710 USA
    }
\end{center}
\vspace{-0.5cm}

\FloatBarrier
\subsection*{S1. Magnetization Measurements and X-Ray Diffractometry of Control and Trilayer Samples}

\vspace{-0.5cm}
\begin{figure}[htb!]
    \centering
    \includegraphics[width=\textwidth]{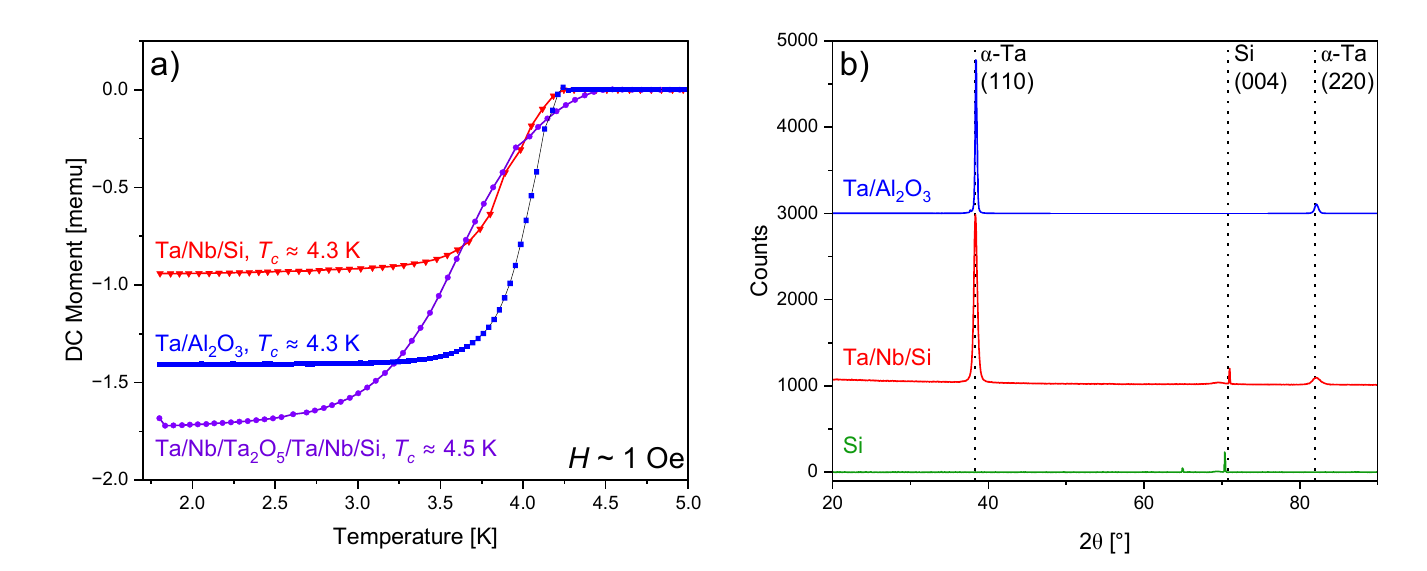}
    \caption{a) Temperature-dependent DC magnetization measurements of both (Si, Al$_2$O$_3$) control Ta films, as well as the trilayer film grown on a Si substrate with Nb seed layers under both the top and bottom Ta layers. The oxide layer for the trilayer was formed through plasma oxidation for 2 hours at 400$\degree$C. All three samples show superconducting transitions. b) XRD spectra of control Ta films grown on on Al$_2$O$_3$ and Si substrates, as well as a bare Si substrate.} 
    \label{S1}
\end{figure}

Magnetization measurements were performed using a Quantum Design MPMS3 SQUID magnetometer to determine the critical temperature ($T_c$) of the tantalum films. Fig.~\ref{S1}a shows the magnetization data of the two control samples: 100-nm Ta film sputtered at room temperature on top of a 6-nm niobium (Nb) seed layer on a silicon (Si) substrate (sample 1) and 150-nm Ta film sputtered on a sapphire (Al$_2$O$_3$) substrate held at 500$\degree$C (sample 2). Both samples exhibit diamagnetic transitions, characteristic of the Meissner effect, with a critical temperature of 4.3 K. Additionally, we collected magnetization data on a trilayer sample grown with Nb seed layers below both the top and bottom Ta layers, with the oxide layer (Ta$_2$O$_5$) formed through heated plasma oxidation at 400$\degree$C for 2 hours (Sample III). This sample also exhibits a diamagnetic transition, with a critical temperature of 4.5 K. The magnitude of the measured moment in the superconducting state at $T=1.8$ K is approximately twice as large as the measured moment of the Ta/Nb/Si control sample at the same temperature and applied magnetic field ($H \sim 1$ Oe). Therefore, both the top and bottom Ta layers contribute to the measured moment in the superconducting state and demonstrate a critical temperature characteristic of $\alpha$-Ta.

The X-ray diffraction (XRD) spectra of the control Ta films grown on both sapphire and Si substrates (Fig.~\ref{S1}b) demonstrate narrow peaks suggestive of high film quality, with prominent $\alpha$-Ta (110) and (220) peaks at $38.3\degree$ and $82\degree$, respectively. To identify the XRD peak associated with the Si substrate, we also performed XRD on a bare Si wafer. We observe a clear Si (004) peak at $69.3\degree$ that is also present for the Ta film grown on Si.

\vspace{-1.0cm}
\FloatBarrier
\subsection*{S2. X-Ray Photoelectron Spectroscopy and Composition Analysis}

\begin{figure}[h!]
    \centering
    \includegraphics[width=0.9\linewidth]{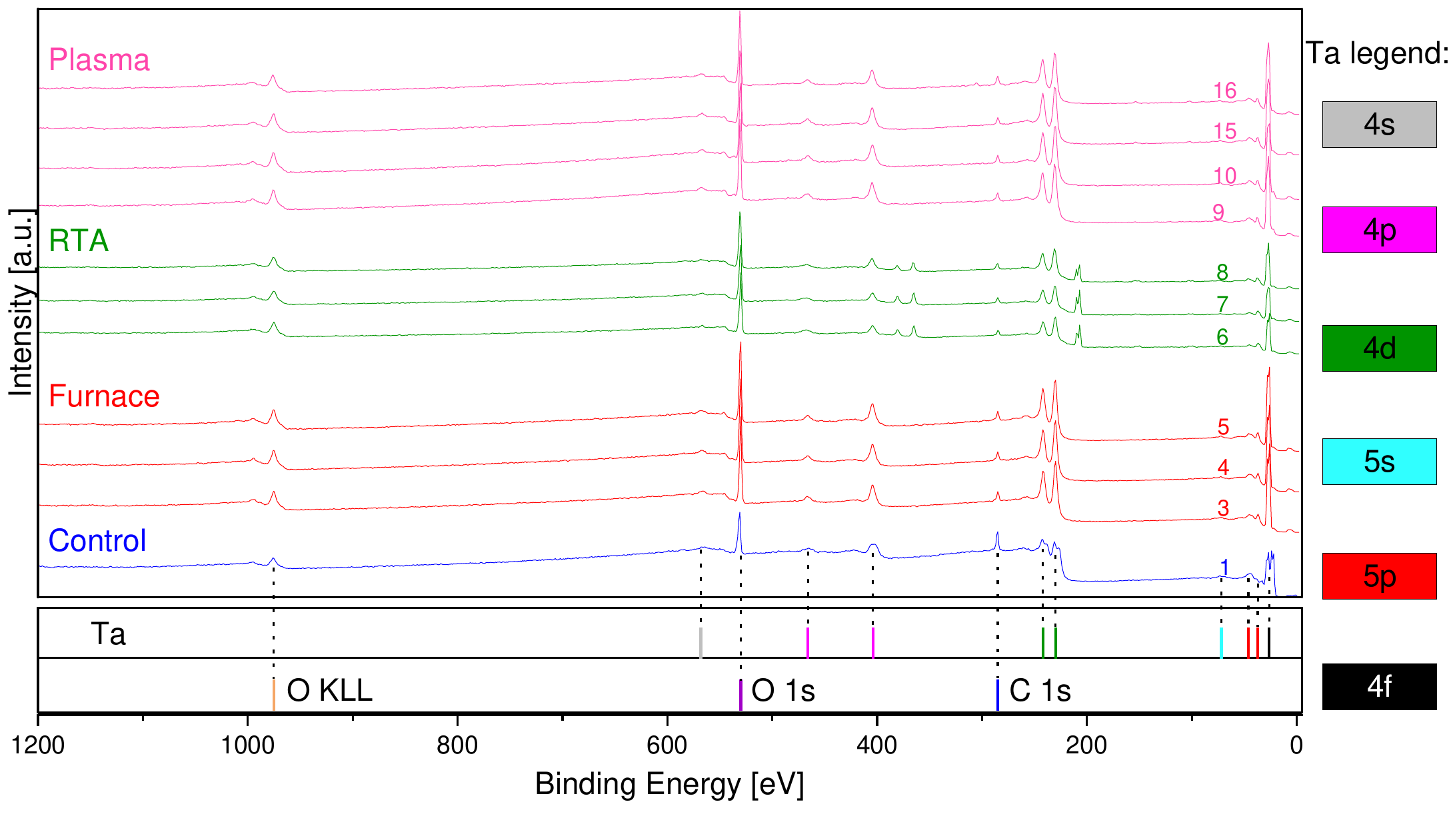}
    \caption{Complete XPS spectra of samples. The bottom panel serves as a key for species observed in these spectra. The legend for Ta peak identification is shown on the right side of the graph. Note that some of the split peaks of Ta cannot be resolved completely in this overview.}
    \label{fig:S6}
\end{figure}

X-Ray Photoelectron Spectroscopy helps assess the surface chemistry, bonding structure and composition of surfaces. The high-resolution spectra of Ta and C can be used to determine what compounds are present on the surface by attributing the binding energy of the peaks to the binding energy of the compound. In the main manuscript, we showed that for samples with oxide layers thicker than 10 nm, the only Ta oxidation state detected near the surface is Ta$^{5+}$, indicating that Ta$_2$O$_5$ is the sole oxide species present. For samples 9 and 10, where the oxide thickness is less than 10 nm ($\sim$7.6 nm), only Ta$^0$ and Ta$^{5+}$ are observed, demonstrating that no significant sub‑oxide species are present. By contrast, the native oxide, as discussed in the main manuscript, contains three Ta oxidation states: 0, 1+, and 5+.

\begin{figure}[h!]
    \centering
    \includegraphics[width=0.8\textwidth]{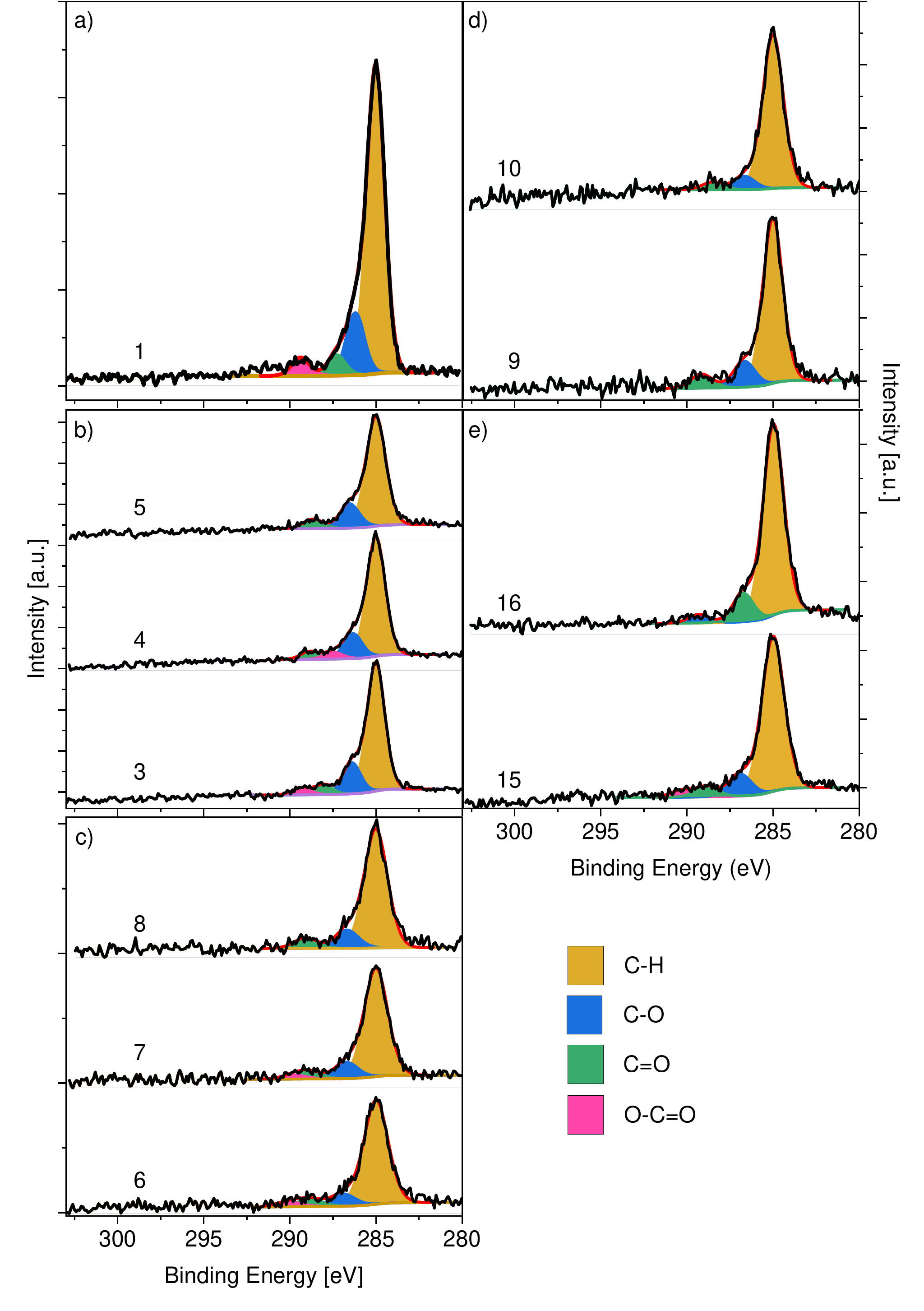}
    \caption{Carbon 1s hi-res spectra for the a) native oxide film (sample 1), b) tube-furnace-oxidized films, c) rapid-thermal-annealed films, d) room temperature plasma-oxidized films, and e) 400 $\degree$C plasma-oxidized films.}
    \label{fig:S7}
\end{figure}

\begin{table*}[!h]
\caption{\label{tab:tableS1} Summary of oxidation process and elemental composition results from XPS spectra. Plasma oxidation was performed with 100 W of RF power biasing the substrate. Refer to Table 1 in the main manuscript for details regarding the layer stack of each sample associated with the IDs in column 1.}

\renewcommand{\arraystretch}{0.8}

\begin{ruledtabular}
\begin{tabular}{cccccccccc}
 & \multicolumn{4}{c}{\cellcolor{gray!20}\textbf{Oxygenation Process}} & \multicolumn{5}{c}{\cellcolor{gray!20}\textbf{Elemental composition (\%)}} \\
\textbf{ID} & \textbf{Method} & \textbf{Temp.} & \textbf{O$_2$} & \textbf{Time} & \textbf{C} & \textbf{O} & \textbf{Ta} & \textbf{Nb} & \textbf{Si} \\
 & & \textbf{[\degree C]} & \textbf{[sccm]} & \textbf{[min]} & \textbf{[1s]} & \textbf{[1s]}& \textbf{[4f]}& \textbf{[3d]}& \textbf{[2p]}\\
\hline
1 & --- & --- & --- & --- & 46& 32.9& 21.1& --- & ---\\
\hline
3 & & & & 10 & 22.6 & 55.8 & 21.6 & --- & ---\\
4 & Tube Furnace & 400 & 20,000 & 30 & 20.6 & 57.4 & 22 & --- & ---\\
5 & & & & 60 & 23.3& 55.3& 21.5& --- & ---\\
\hline
6 & & & & 1 & 16.3& 56& 13.4& 7.4& 6.9\\
7 & RTA & 700 & 5 & 5 & 18.1& 54.9& 12.5& 8.1& 6.4\\
8 & & & & 10 & 19.7& 55.3& 15& 5.7& 4.4 \\
\hline
9 & & 21.3 & & 60 & 21.8& 56.8& 21.4& --- & --- \\
10 & Plasma & 21.3 & 20 & 120 & 19.2 & 57 & 23.9 & --- & --- \\
15 & & 400 & & 60 & 14.4& 59.2& 21.8& ---& 4.6\\
16 & & 400 & & 120 & 19.7& 56.6& 20.3& ---& 3.5\\
\end{tabular}
\end{ruledtabular}

\renewcommand{\arraystretch}{1.0} 
\end{table*}

For carbon, multiple types of bonds have been found on the surfaces of different samples.  Fig.~\ref{fig:S7} (a-e) shows the peak-fit analysis for the C 1s hi-res spectra, for different types of oxidation. All films contain hydrocarbons (C-C), ethers/alcohols (C-O), and carbonyls (C=O) on the surface, while some samples also contain carboxylic acids (O-C=O).

\FloatBarrier
\subsection*{S3. Optical and Scanning Electron Micrographs of Oxidized Ta Films}

\begin{figure}[h!]
    \centering
    \includegraphics[width=1\textwidth]{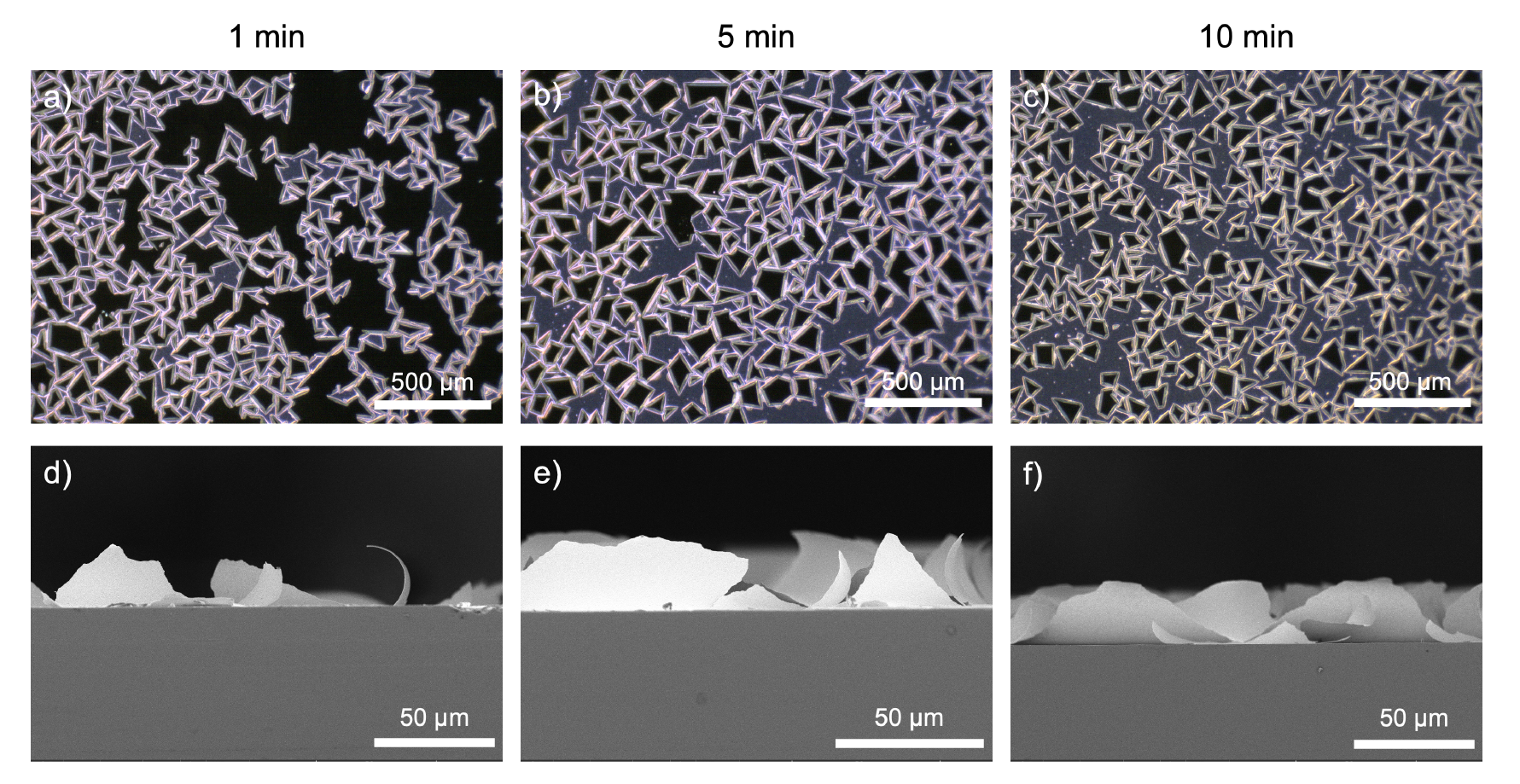}
    \caption{Optical microscope images of sample surfaces (a-c) and cross-sectional SEM images (d-f) after oxidation of Ta films through rapid thermal annealing at 700\degree C for 1, 5, and 10 minutes. Significant film delamination can be observed. Dark regions in the optical microscope images correspond to the Si substrate.}
    \label{RTA}
\end{figure}

\begin{figure}[h!]
    \centering
    \includegraphics[width=1\textwidth]{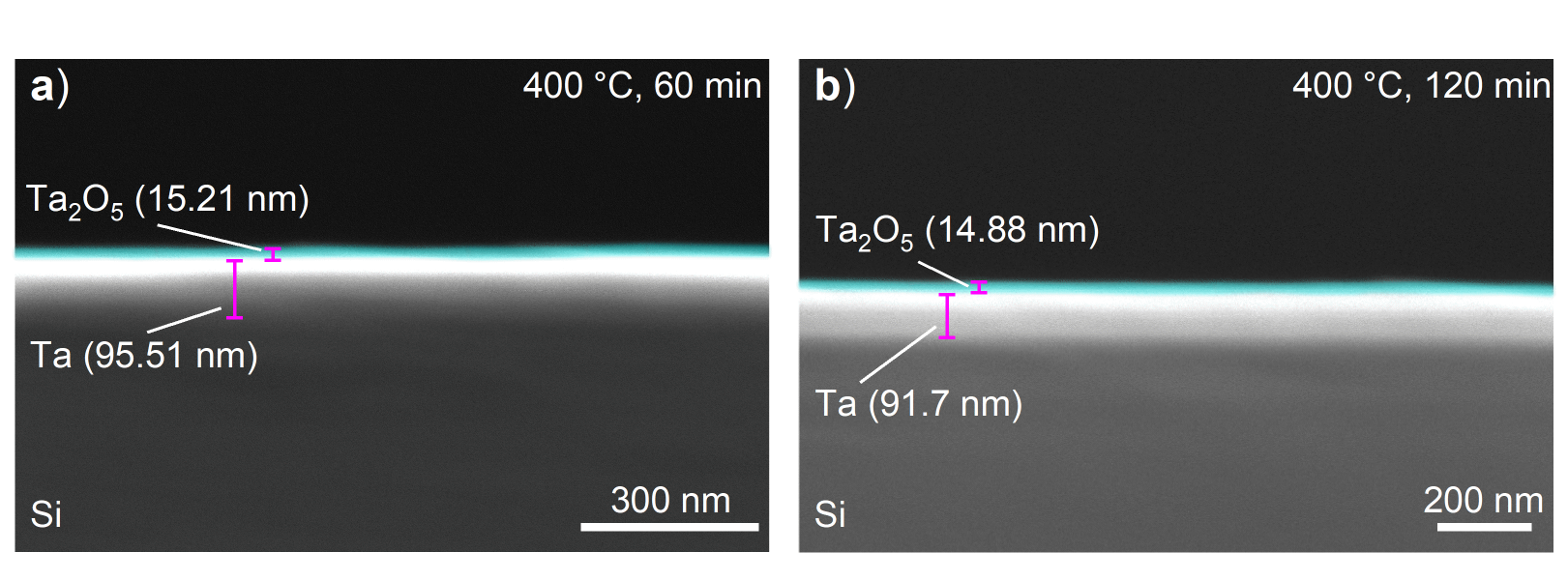}
    \caption{SEM images showing cross sections of tantalum films that underwent plasma oxidation at 400 $\degree$C for a) 60 min and b) 120 min. The Ta$_2$O$_5$ layer is highlighted in false color, and the thickness values of the Ta and Ta$_2$O$_5$ layers extracted from XRR measurements are labeled. The tantalum film on both samples was initially deposited at room temperature on a 6-nm Nb seed layer, which is more difficult to distinguish from the Si substrate due to insufficient contrast.}
    \label{SEM}
\end{figure}

Scanning electron microscopy (SEM) images of the films oxidized using an oxygen plasma and rapid thermal annealing were collected using a ThermoFisher Apreo SEM at working distances of 3-5 mm. Cross sections of plasma-oxidized Ta films are shown in Fig.~\ref{SEM}. The observed thicknesses of the Ta and Ta$_2$O$_5$ layers in SEM correspond well to the thickness values extracted from X-ray reflectometry (XRR) measurements. Fig.~\ref{RTA} presents optical surface images, as well as the cross-sectional SEM images, of Ta films after oxidation by rapid thermal annealing. All RTA oxidized films show clear delamination from the Si substrate, involving both the Ta and Nb seed layers.

\FloatBarrier
\subsection*{S4. Interstitial Oxygen Diffusion in $a$-Ta$_2$O$_5$}

Excess oxygen in fully oxidized amorphous Ta$_2$O$_5$ takes the form of a peroxy species (O$_2^{2-}$), i.e., a neutral oxygen O$^0$ atom bound to a lattice O$^{2-}$ ion in a dumbbell configuration. To determine the oxygen diffusion barrier, we first generated a model $a$-Ta$_2$O$_5$ structure, represented by a Ta$_{36}$O$_{90}$ supercell. The initial structure was obtained by randomly distributing Ta and O atoms over the supercell volume. Then the total energy of this structure was minimized with respect to the atomic coordinates and supercell parameters; the resulting structure was further modified by applying atomic displacements and followed by one more energy minimization step. The displacement directions were selected at random, and the magnitude of the displacements was varied (also at random) between 1 and 3 \AA. The lowest energy configuration was selected for further calculations.

\begin{figure}[!h]
    \centering
    \includegraphics[width=0.6\textwidth]{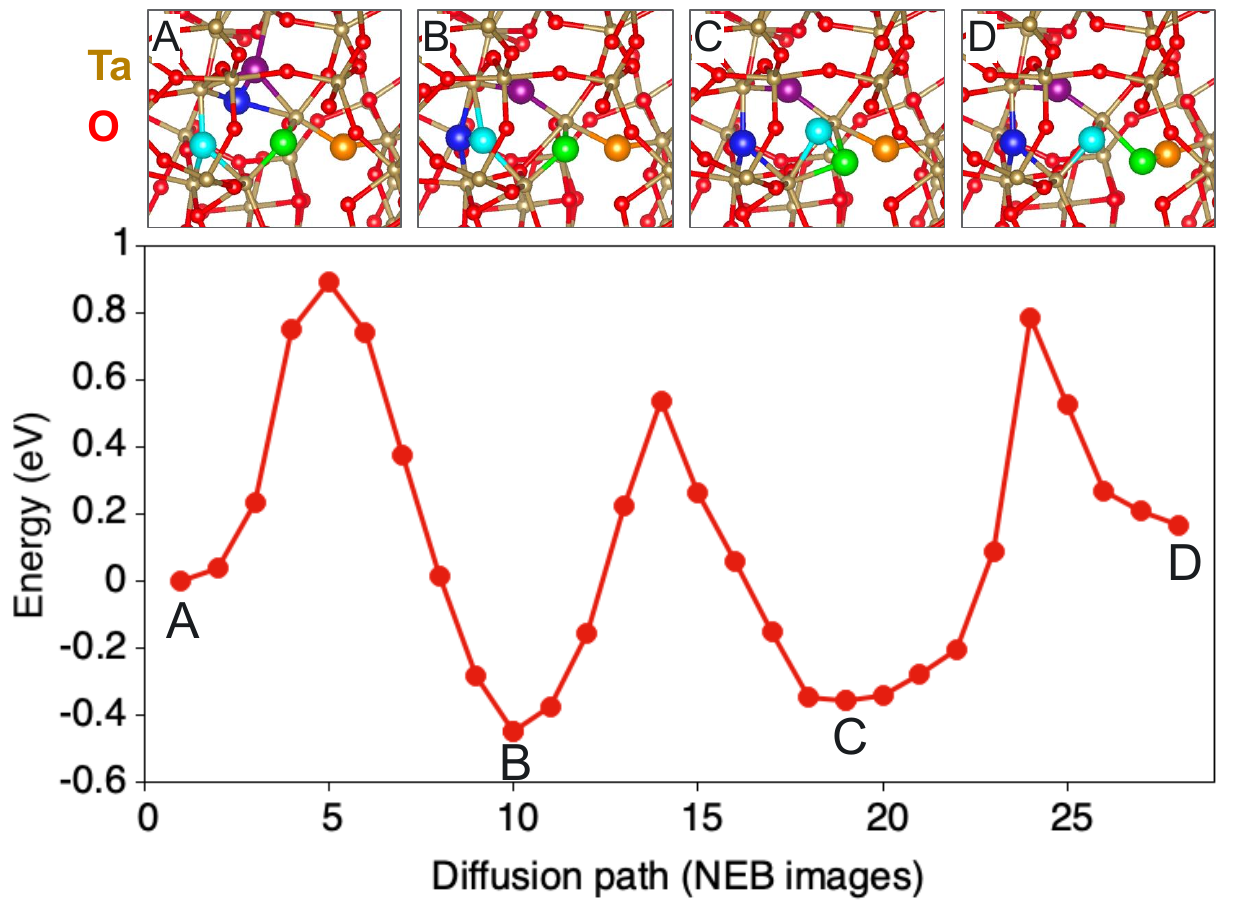}
    \caption{Ball-and-stick representations of the local atomic configurations of the excess oxygen atom along its diffusion path in $a$-Ta$_2$O$_5$ (top) and the corresponding potential energy profile (bottom). Large colored spheres indicate five oxygen atoms participating in the considered interstitialcy oxygen diffusion. In all cases, the excess O$^0$ atom binds to a lattice O$^{2-}$, thus forming a dumbbell O$_2^{2-}$ configuration. The diffusion of O$^0$ proceeds by breaking one dumbbell configuration, displacing the atom toward a neighboring O$^{2-}$ and forming a new dumbbell configuration, as indicated with the rearrangement of the large colored spheres along the A--B--C--D path.}
    \label{fig:DFT-O-in-Ta2O5}
\end{figure}

The stability of point defects in an amorphous lattice depends on the details of their local environment. We analyzed the effect of this variability on the stability of excess O$^0$ by considering four configurations and calculating six diffusion barriers between these configurations (see Fig.~\ref{fig:DFT-O-in-Ta2O5}). Here, the O species participating in the diffusion process are shown with large colored spheres (purple, blue, cyan, green, and orange). The initial configuration (A) is formed by the purple-blue dumbbell O$_2^{2-}$ ion. Transition to configuration B corresponds to breaking the purple-blue bond and forming the blue-cyan bond, and so on, as shown in panels C (cyan-green) and D (green-orange). Our sampling indicates that the O diffusion barrier varies between 0.6 and 1.2 eV, depending on the local configuration, which points to the possibility of a facile O diffusion at elevated temperatures. These calculations also suggest that the formation of interstitial oxygen is thermodynamically unfavorable relative to the gas-phase O$_2$ molecule by 0.8--1.4 eV. Thus, we conclude that forming an appreciable concentration of these ions is possible only under high oxygen chemical potential conditions, such as high O$_2$ pressure and elevated temperature. However, once formed and in the presence of high concentration gradients, these species provide a potent channel for enhanced Ta oxidation that is not available at normal conditions.

\FloatBarrier
\subsection*{S5. Scanning Transmission Electron Microscopy}
\begin{figure}[!htb]
    \centering
    \includegraphics[width=0.6\textwidth]{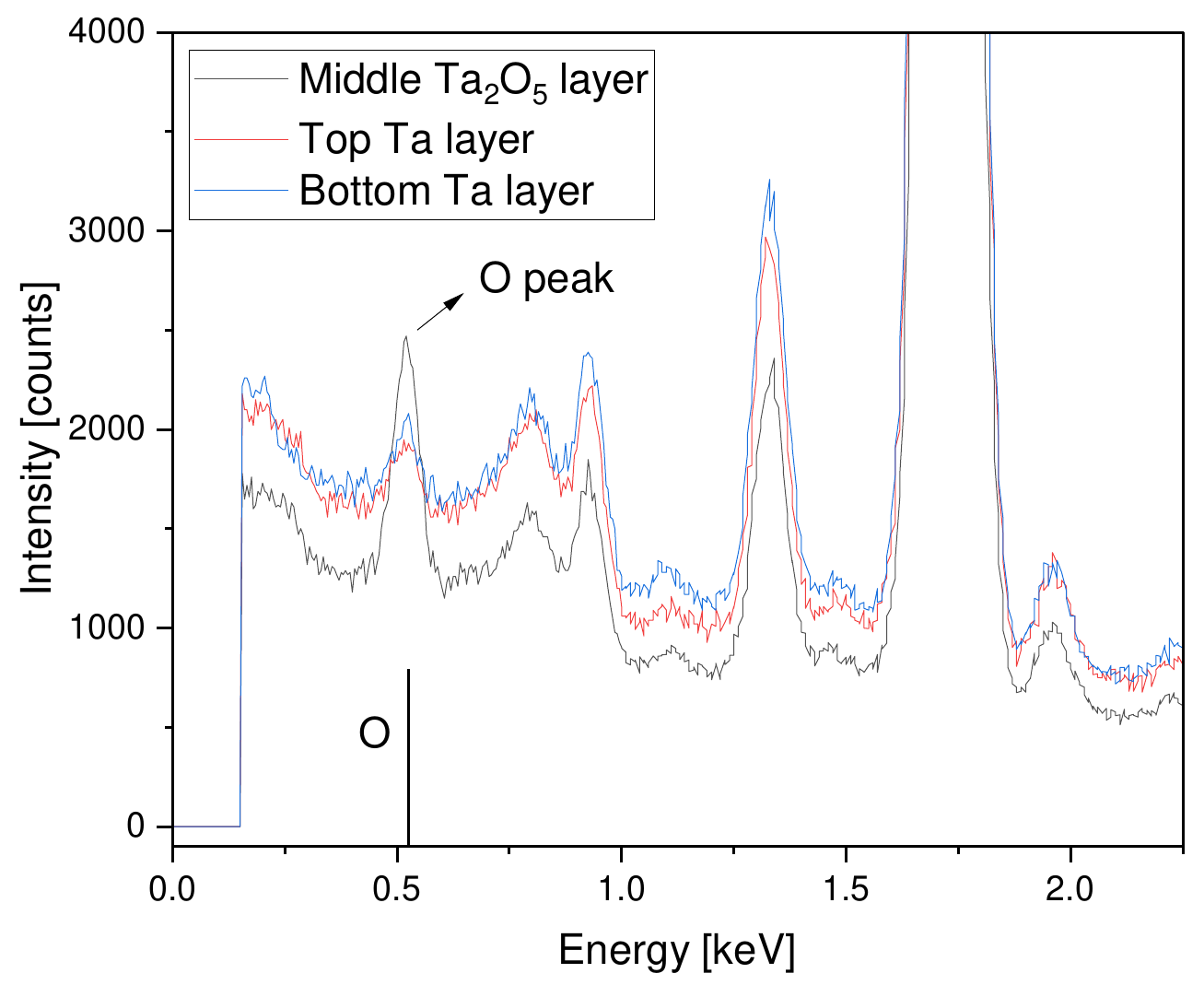}
    \caption{EDX spectra of the Ta/Ta$_2$O$_5$/Ta trilayer \RomanNumeral{7}, associated with Fig.~6 in the main manuscript. The red, black, and blue curves correspond to the top Ta electrode, the Ta$_2$O$_5$ barrier, and the bottom Ta electrode, respectively. An oxygen peak at 0.525~keV appears in all three layers and is predominately attributed to bulk contamination in the lamellae TEM specimen.}
    \label{fig:EDX}
\end{figure}

The energy-dispersive X-ray (EDX) spectroscopy results shown in Fig.~\ref{fig:EDX} quantify oxygen distribution across the Ta/Ta$_2$O$_5$/Ta stack. Spectra were acquired from three regions: top Ta electrode, Ta$_2$O$_5$ barrier, and bottom Ta electrode. The oxygen peak at 0.525 keV shows maximum intensity ($\sim$ 2500 counts) in the Ta$_2$O$_5$ barrier as expected.

\begin{figure}[!htb]
    \centering
    \includegraphics[width=1\linewidth]{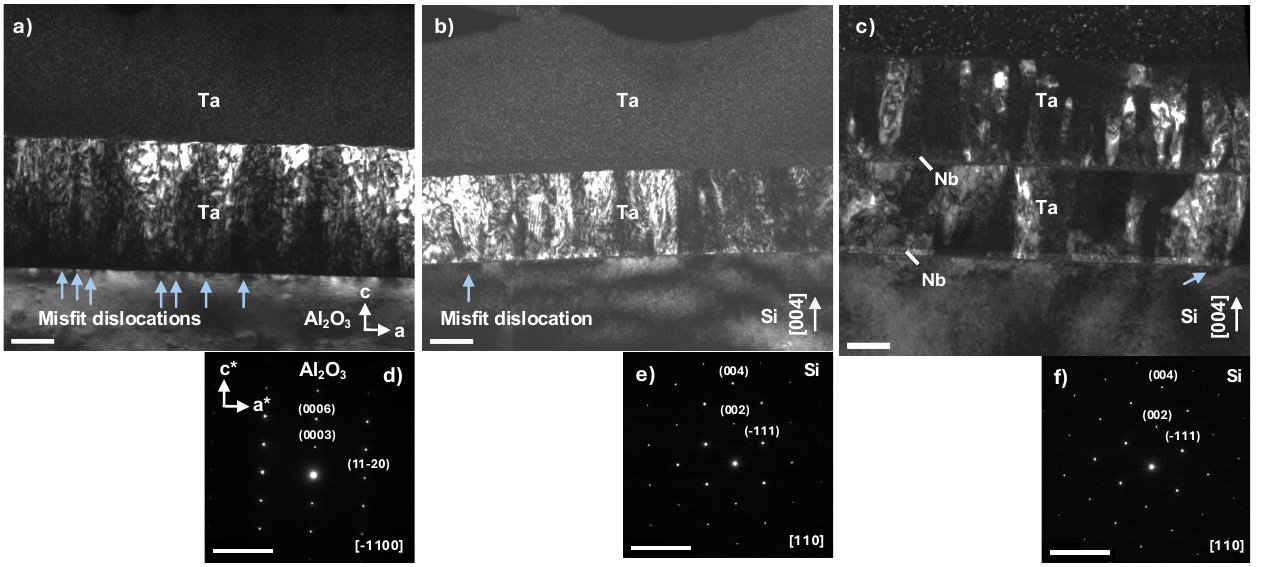}
    \caption{Cross-sectional TEM dark-field images (a-c) and corresponding selected-area electron diffraction (SAED) patterns (d-f) highlighting the substrate–material interfaces of (a) Ta/Ta$_2$O$_5$/Ta on sapphire (trilayer \RomanNumeral{7}), (b) Ta/Ta$_2$O$_5$/Ta/ on Nb/Si (sample \RomanNumeral{5}), and (c) Ta/Nb/Ta$_2$O$_5$/Ta on Nb/Si (sample \RomanNumeral{1}). For the SAED images, the zone axes is noted in the lower left corner.}
    \label{fig:SMDFTEM}
\end{figure}

\begin{figure}[!htb]
    \centering
    \includegraphics[width=1\linewidth]{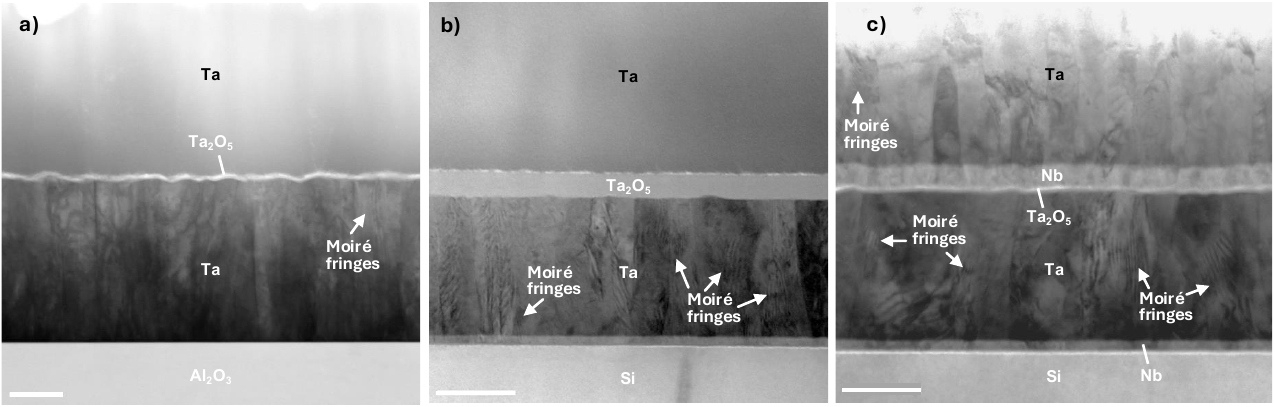}
    \caption{Cross-sectional bright-field STEM images of Ta/Ta$_2$O$_5$/Ta trilayer stacks for (a) trilayer \RomanNumeral{7} (Ta/Ta$_2$O$_5$/Ta on sapphire), (b) trilayer \RomanNumeral{5} (Ta/Ta$_2$O$_5$/Ta on Nb/Si), and (c) trilayer \RomanNumeral{1} (Ta/Nb/Ta$_2$O$_5$/Ta on Nb/Si). }
    \label{fig:SMBFTEM}
\end{figure}

Fig.~\ref{fig:SMDFTEM} shows the dark-field cross-sectional TEM images for three trilayers: Ta/Ta$_2$O$_5$/Ta (trilayer \RomanNumeral{7}) on sapphire,  Ta/Ta$_2$O$_5$/Ta on Nb/Si (trilayer \RomanNumeral{5}), and Ta/Nb/Ta$_2$O$_5$/Ta on Nb/Si (trilayer \RomanNumeral{1}). A high density of misfit dislocations was observed at the Ta/sapphire interface in trilayer \RomanNumeral{7}, indicated by blue arrows in Fig.~\ref{fig:SMBFTEM}(a), reflecting a large lattice mismatch between $\alpha$-Ta and Al$_2$O$_3$. These misfit dislocations facilitate the nucleation and propagation of threading dislocations within the Ta layers. In contrast, only a few misfit dislocations were observed at the Si/Nb/Ta interfaces for trilayers \RomanNumeral{5} and \RomanNumeral{1}, indicating improved lattice accommodation on silicon substrates, mediated by the Nb seed layer. The dark-field image of trilayer \RomanNumeral{7} was formed using the sapphire (0006) reflection, whereas the Si (004) reflection was used for trilayers \RomanNumeral{5} and \RomanNumeral{1} 

\begin{figure}[!htb]
    \centering
    \includegraphics[width=1\linewidth]{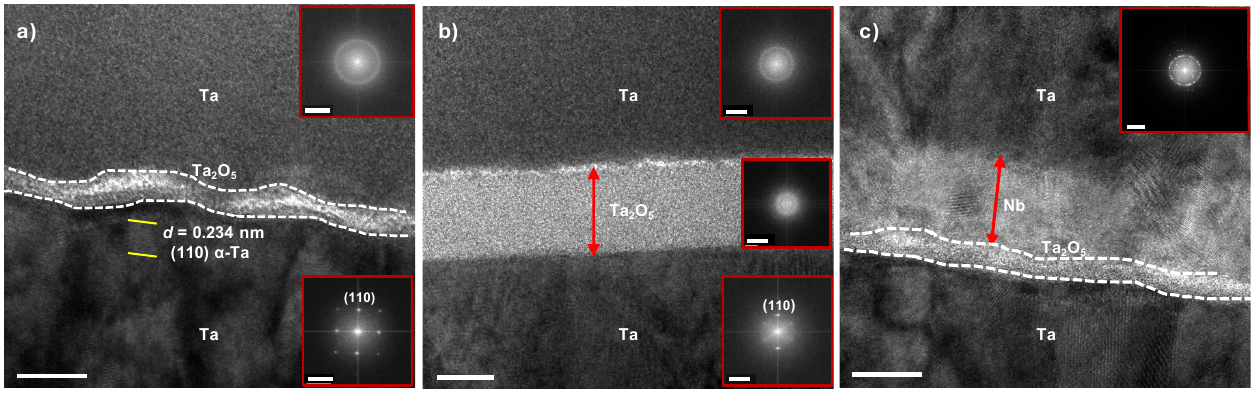}
    \caption{Cross-sectional high-resolution transmission electron microscopy images of Ta/Ta$_2$O$_5$/Ta trilayer stacks acquired for (a) Ta/Ta$_2$O$_5$/Ta on sapphire (trilayer \RomanNumeral{7}), (b) Ta/Ta$_2$O$_5$/Ta on Nb/Si (trilayer \RomanNumeral{5}), and (c) Ta/Nb/Ta$_2$O$_5$/Ta on Nb/Si (trilayer \RomanNumeral{1}), shown from left to right. The insets show FFT patterns for each layer.}
    \label{fig:SMHRTEM}
\end{figure}

Fig.~\ref{fig:SMHRTEM} shows cross-sectional high-resolution transmission electron microscopy (HRTEM) images of the same set of trilayers shown in Fig.~\ref{fig:SMBFTEM}. In all samples, an amorphous Ta$_2$O$_5$ barrier layer is clearly observed between the top and bottom Ta layers, as indicated by the dashed white lines. Fast Fourier transform (FFT) patterns, shown in the insets in Fig.~\ref{fig:SMHRTEM} (a–b), confirm the amorphous nature of the top Ta layer and the crystalline character of the bottom Ta layer in trilayers \RomanNumeral{7} and \RomanNumeral{5}. In trilayer \RomanNumeral{7}, the marked lattice fringes correspond to the (110) plane of $\alpha$-Ta, with an interplanar spacing of approximately 0.234 nm. Both the top and bottom Ta layers are crystalline in trilayer \RomanNumeral{1}. The FFT acquired from the Ta/Ta$_2$O$_5$/Nb/Ta interfacial region in \RomanNumeral{1} reveals a polycrystalline structure, indicating the presence of grains with multiple orientations.

Lastly, Fig.~\ref{fig:SMBFTEM} presents bright-field TEM images of three of the trilayers. For all, we observe Moir\'e fringes, which originate from the overlap of misoriented crystalline Ta grains. Images of trilayer \RomanNumeral{7} in Fig.~\ref{fig:SMBFTEM}(a) reveal that Moir\'e fringes are minimal, consistent with columnar Ta grain growth predominantly aligned along the c-axis of the sapphire substrate. In contrast, trilayers \RomanNumeral{5} and \RomanNumeral{1} grown on Si substrates exhibit more pronounced Moir\'e fringes within the Ta layers, indicating the presence of misoriented grains. In trilayer \RomanNumeral{1}, the columnar grains are not perfectly aligned, resulting in a higher density of Moir\'e fringes compared to the other samples.